\documentclass[12pt]{article}
\usepackage{amsmath,amssymb,epsfig}
\usepackage{epsfig}
\renewcommand{\theequation}{\thesection.\arabic{equation}}
\newlength{\extraspace}
\newlength{\extraspaces}
\newcommand{\be}{\begin{equation}
\addtolength{\abovedisplayskip}{\extraspaces}
\addtolength{\belowdisplayskip}{\extraspaces}
\addtolength{\abovedisplayshortskip}{\extraspace}
\addtolength{\belowdisplayshortskip}{\extraspace}}
\newcommand{\ee}{\end{equation}}
 
\newcommand{\ba}{\begin{eqnarray}
\addtolength{\abovedisplayskip}{\extraspaces}
\addtolength{\belowdisplayskip}{\extraspaces}
\addtolength{\abovedisplayshortskip}{\extraspace}
\addtolength{\belowdisplayshortskip}{\extraspace}}
\newcommand{\ea}{\end{eqnarray}}

\newcommand{\bas}{\begin{eqnarray*}
\addtolength{\abovedisplayskip}{\extraspaces}
\addtolength{\belowdisplayskip}{\extraspaces}
\addtolength{\abovedisplayshortskip}{\extraspace}
\addtolength{\belowdisplayshortskip}{\extraspace}}
\newcommand{\eas}{\end{eqnarray*}}
 
\newcounter{subequation}[equation]
\makeatletter

\expandafter\let\expandafter
\reset@font\csname reset@font\endcsname

\def\subeqnarray{\arraycolsep1pt
    \def\@eqnnum\stepcounter##1{\stepcounter{subequation}%
        {\reset@font\rm(\theequation\alph{subequation})}}
\jot5mm     \eqnarray}

\def\subarray{\arraycolsep1pt
    \def\@eqnnum\stepcounter##1{\stepcounter{subequation}%
        {\reset@font\rm(\alph{subequation})}}
\jot5mm     \eqnarray}

\makeatother
\newcommand{\newappendix}[1]{
\vspace{15mm}
\pagebreak[3]
\addtocounter{section}{1}
\setcounter{equation}{0}
\setcounter{subsection}{0}
\renewcommand{\theequation}{\Alph{section}.\arabic{equation}}
\begin{flushleft}
{\large\bf Appendix \Alph{section}: #1}
\end{flushleft}
\nopagebreak
\medskip
\nopagebreak}

\newcommand{\newsection}[1]{
\vspace{15mm}
\pagebreak[3]
\addtocounter{section}{1}
\setcounter{equation}{0}
\setcounter{subsection}{0}
\begin{flushleft}
{\large\bf \thesection. #1}
\end{flushleft}
\nopagebreak
\medskip
\nopagebreak}
 
\newcommand{\newsubsection}[1]{
\vspace{1cm}
\pagebreak[3]
 
\addtocounter{subsection}{1}
\noindent{ \bf \thesection.\arabic{subsection} #1}
\nopagebreak
\vspace{2mm}
\nopagebreak}

\newcommand{\Z}{\mathbb{Z}}
\newcommand{\R}{\mathbb{R}}

\newcommand{\bra}{\langle}
\newcommand{\ket}{\rangle}
\newcommand{\ra}{\rightarrow}

\newcommand{\rra}{\ \longrightarrow \ }

\newcommand{\is}{ &\! =\! & }
\newcommand{\nonum}{\nonumber \\[1.5mm]}
\newcommand{\sspace}{\makebox[1cm]{ }}

\newcommand{\nspace}{\!\!\!\!\!\!\!\!\!\!}

\newcommand{\Tr}{{\rm Tr}}

\newcommand{\lb}{\lambda}
\newcommand{\om}{\omega}
\newcommand{\sh}{{\rm sh}}

\newcommand{\rme}{{\rm e}}
\newcommand{\rmO}{{\rm O}}
\newcommand{\rmd}{{\rm d}}
\newcommand{\dd}{{\partial}}

\newcommand{\cN}{{\cal N}}

\begin{document}

%%%%%%%%%%%%%%%%%%%%%%%%%%%%%%%%%%%%%%%%%%%%%%%%%%%%%%%%%%%%%%%%%%%%%%%
\begin{titlepage}

%footnotesymbols others than numbers
\renewcommand{\thefootnote}{\fnsymbol{footnote}}
\begin{flushright}
MPP-2007-75
\end{flushright}
\mbox{}
%\vspace{-5mm}

\begin{center}
\mbox{{\Large \bf Noncompact sigma-models:}}\\[4mm]
\mbox{{\Large \bf Large $N$ expansion and thermodynamic limit}}\\[4mm]
\vspace{1.3cm}

{{\sc A. Duncan, M.~Niedermaier\footnote{Membre du CNRS}, P. Weisz}}
\\[4mm]
{\small\sl Department of Physics}\\
{\small\sl 100 Allen Hall, University of Pittsburgh} \\
{\small\sl Pittsburgh, PA 15260, USA}
\\[3mm]
{\small\sl Laboratoire de Mathematiques et Physique Theorique}\\
{\small\sl CNRS/UMR 6083, Universit\'{e} de Tours}\\
{\small\sl Parc de Grandmont, 37200 Tours, France}
\\[3mm]
{\small\sl Max-Planck-Institut f\"{u}r Physik}\\
{\small\sl F\"ohringer Ring 6}\\
{\small\sl 80805 M\"unchen, Germany}

\vspace{1.2cm}
\end{center}

\begin{quote}
Noncompact ${\rm SO}(1,N)$ sigma-models are studied in terms of 
their large $N$ expansion in a lattice formulation in dimensions 
$d \geq 2$. Explicit results for the spin and current two-point 
functions as well as for the Binder cumulant are presented to 
next to leading order on a finite lattice. The dynamically 
generated gap is negative and serves as a coupling-dependent 
infrared regulator which vanishes in the limit of infinite lattice 
size. The cancellation of infrared divergences in invariant correlation 
functions in this limit is nontrivial and is in $d=2$ demonstrated by 
explicit computation for the above quantities. For the Binder 
cumulant the thermodynamic limit is finite and is given by 
$2/(N\!+\!1)$ in the order considered. Monte Carlo simulations
suggest that the remainder is small or zero. The potential 
implications for ``criticality'' and ``triviality'' of the theories 
in the ${\rm SO}(1,N)$ invariant sector are discussed.  
\end{quote} 
\vfill

\setcounter{footnote}{0}
\end{titlepage}

%%%%%%%%%%%%%%%%%%%%%%%%%%%%%%%%%%%%%%%%%%%%%%%%%%%%%%%%%%%%%%%%%%%%%%%%%%%%

\newsection{Introduction} 

In quantum field theories with nonabelian symmetries and 
dynamical mass generation the large $N$ expansion often 
provides a qualitatively correct and quantitatively reasonable 
description of the physics of the systems. Specifically in 
sigma-models with a compact global symmetry group the 
expansion is known to be an asymptotic expansion \cite{kupi} 
and when slightly ad hoc applied to low orders at fixed 
small $N$ sometimes gives surprisingly accurate results, see 
e.g.~\cite{ffgr,campostrini} for the renormalized coupling. In 
a lattice formulation one starts off on a finite lattice, the 
associated `finite volume' mass gap then is 
uniformly bounded away from zero, and in the 
large $N$ series for invariant correlation functions the 
limit of infinite lattice size (also called the 
thermodynamic limit) can safely be taken termwise.

A study of the large $N$ expansion in sigma-models 
having a noncompact ${\rm SO}(1,N)$ internal symmetry 
group has been initiated in \cite{DNS, EMP}. A large 
value of $N$ in this case is also physically relevant for the 
granular limit of random hamiltonians describing 
disordered electrons with $N$ orbitals per site \cite{disorder}. 
In a lattice formulation of the ${\rm SO}(1,N)$ sigma-models 
again a gap is dynamically generated in the large $N$ expansion, 
which is however negative and vanishes as the size 
of the lattice goes to infinity. Effectively the gap now  
acts as a subtle, coupling-dependent, infrared regulator and 
the technical problem consists in studying the `coordinated' limit 
$V \ra \infty$ of lattice sums of the form 
$\frac{1}{V^n} \sum_{k_1, \ldots, k_n} f_V(k_1, \ldots, k_n)$, 
where $f_V$ carries an explicit $V$-dependence via the 
gap. The sums associated with individual Feynman diagrams 
of the large $N$ expansion will typically diverge in the limit. 
The issue whether or not in the combinations entering 
invariant correlation functions the infrared divergences 
cancel is analogous to the one encountered in the perturbation 
theory of compact sigma-models \cite{David,Elitzur} and is the 
subject of the present paper. Since this issue 
is most critical in the two-dimensional systems we 
examine the limit specifically in this case, although our 
finite volume results are valid in all dimensions $d \geq 2$.  
We compute a number of physically interesting quantities to 
leading and subleading order and show that they indeed do have 
a well-defined thermodynamic limit. Concretely we consider the 
spin two-point function, the two-point function of the Noether 
current, and the Binder cumulant. 

The Binder cumulant $U$ is defined in terms of the zero momentum 
limit of the connected four-point function. In massive
scalar field theories it serves to define an intrinsic 
measure of the interaction strength and has been used 
to explore ``triviality'' issues. In a massless theory, 
like the systems considered here, there is no obvious 
reason why $U$ should have a finite thermodynamic limit.  
Somewhat surprisingly we find that $U$ does have a finite 
and nonzero limit to leading and subleading order,
which is moreover independent of $\lb$ and given by 
$2/(N\!+\!1)$. Supported also by Monte-Carlo simulations 
we conjecture that the infinite volume limit of the 
exact $U$ is also very close to $2/(N\!+\!1)$. The potential
implications for ``criticality'' and ``triviality'' in the
${\rm SO}(1,N)$ invariant sector of the theory will be 
discussed in the conclusions.  

The rest of the article is organized as follows. In the 
next section we review a result from a previous paper 
\cite{EMP} which on a finite lattice allows one to do 
large $N$ computations in the simpler compact models and 
then transfer the results to the noncompact ones 
via a ``large $N$ correspondence''. This correspondence has 
an interesting interplay with the Schwinger-Dyson equations  
and instead of going through the (fairly routine) diagrammatic 
computations we merely present the results as solutions of 
the large $N$ expanded Schwinger-Dyson equations with the 
correct `initial' data. Expressions for the two- and four 
point functions to leading and subleading 
order are given (valid on a finite lattice in all dimensions
$d \geq 2$), from which also the two-point function of 
the Noether current and the Binder cumulant can be  
obtained in the same order. The thermodynamic limit
in $d=2$ of the local quantities and of the Binder cumulant 
are studied in Sections~4 and 5, respectively.

%%%%%%%%%%%%%%%%%%%%%%%%%%%%%%%%%%%%%%%%%%%%%%%%%%%%%%%%%%%%%%%%%%%%%%%%%%%%

\newsection{Large $N$ expansions of compact and non\-compact models} 

In compact sigma-models the large $N$ expansion is a saddle 
point expansion based on a generating functional obtained 
by `dualizing' the spins, i.e.~by imposing the constraint 
via a Lagrange multiplier field and performing the 
Gaussians.  The counterpart of this duality 
transformation is somewhat ill-defined in the noncompact models.
The large $N$ expansion can nevertheless be justified 
and on a {\it finite lattice} the expansion coefficients 
for invariant correlation functions can be inferred 
from those in the compact model \cite{EMP}. This 
``large $N$ correspondence'' allows one to do computations 
in the compact model, where no gauge gauge-fixing is 
required, and the familiar framework can be used. 
Here we briefly summarize the correspondence 
and present explicit results for two and four-point 
functions in Section~3. The results of Sections~2 and 3 
are valid in all dimensions $d \geq 2$.

%%%%%%%%%%%%%%%%%%%%%%%%%%%%%%%%%%%%%%%%%%%%%%%%%%%%%%%%%%%%%%%%%%%%%%%

\newsubsection{Definitions} 

Here we recall the notation and the definitions for the invariant 
correlation functions considered and their generating functionals.
We consider the ${\rm SO}(N+1)$ spherical and the ${\rm SO}(1,N)$ 
hyperbolic sigma-models in two dimensions with standard lattice action, 
defined on a hypercubic lattice $\Lambda \subset \Z^d$ of volume 
$V = |\Lambda| = L^d$. The dynamical variables (``spins'') 
will be denoted by $n_x^a$, $x \in \Lambda$, $a =0, \ldots, N$, in both 
cases, and periodic boundary conditions are assumed throughout 
$n_{x + L\hat{\mu}} = n_x$. The constraint is $n\cdot n =1$ in 
both cases, but with different `dot' products; namely $a \cdot b := 
a^0 b^0 + a^1 b^1 + \ldots + a^N b^N$ in the compact 
model, and $a \cdot b := a^0 b^0 - a^1 b^1 - \ldots - a^N b^N$ 
in the noncompact model.
Clearly $S^N = \{ n \in \R^{N+1}\,| \,n\cdot n =1\}$ is the 
$N$-sphere and $H^N = \{ n \in \R^{1,N}\,|\, n \cdot n =1,\, n^0 >0\}$ 
is the upper half of the two-sheeted $N$-dimensional hyperboloid. 
The lattice actions are 
\be 
S_{\pm} = \mp \beta \sum_{x,\mu} (n_x \cdot n_{x+\hat{\mu}} -1) = 
\mp \frac{\beta}{2} \sum_x n_x \cdot (\Delta n)_x \geq 0\,,
\label{c2}
\end{equation}
where the upper sign refers to the compact model and the lower 
sign to the noncompact model. The laplacian is 
$\Delta_{xy} = - \sum_{\mu} [2 \delta_{x,y} - \delta_{x,y + \hat{\mu}} 
- \delta_{x, y - \hat{\mu}}]$, as usual. We write 
\ba 
\rmd\Omega_+(n) \is \rmd^{N+1}n\,\delta(n\cdot n -1)\,,
\nonum
\rmd\Omega_-(n) \is 2 \rmd^{N+1}n\, \delta(n\cdot n -1) \theta(n^0)\,,
\label{c3}
\ea
for the invariant measure on $S^N$ and $H^N$, respectively.
Further $\delta_{\pm}(n,n')$ is the 
invariant point measure on $S^N$, $H^N$, and 
$n^{\uparrow} = (1,0,\ldots,0)$. Note that the 
measure $\rmd\Omega_+(n)$ is normalized while $H^N$ has 
infinite volume. 

In the compact model we consider the generating functional,
\be
\nspace \exp W_+[H] = \cN \!\int\! \prod_{x}\rmd\Omega_+(n_x) \exp\Big\{
\!- S_+ + \frac{1}{2} \sum_{x,y} H_{xy}( n_x \cdot n_y -1) \Big\}\,,
\label{wc1}
\end{equation}
where $H_{xy} \geq 0$ is a source field and the normalization $\cN$ is such 
that $W[0] = 0$. 
For the noncompact model we consider the generating functional
\ba 
\nspace \exp W_-[H]\! \is \!\cN \!\int\! \prod_{x}\rmd\Omega_-(n_x) 
\delta_-(n_{x_0},n^{\uparrow})\,
\exp\Big\{\!
- S_- + \frac{1}{2} \sum_{x,y} H_{xy}( n_x \cdot n_y -1) \Big\}\,,
\label{wnc}
\ea 
where now $H_{xy} <0$ sources give damping exponentials, and
one spin at site $x_0$ is fixed in order to make the 
generating functional well defined. 

Connected $2r$ point functions are defined by 
\ba 
&& W_{\pm}[H] = \sum_{r \geq 1} \frac{1}{r! \, 2^r}\, 
W_{\pm,r}(x_1,y_1;\ldots;x_r, y_r) \, H_{x_1 y_1} \ldots 
H_{x_r y_r} \,,
\nonum
&& W_{\pm,r}(x_1,y_1;\ldots;x_r, y_r) := h_{x_1 y_1} \ldots h_{x_r y_r}
W_{\pm}[H]\Big|_{H=0}\,,\quad h_{xy} := \frac{\delta}{\delta H_{xy}}\,.
\label{c4}
\ea
In particular $W_{\pm,1}(x,y) = \bra n_x \cdot n_y\ket_{\pm} -1$, 
$W_{\pm,2}(x_1,y_1;x_2, y_2) := \bra n_{x_1} \cdot n_{y_1}  n_{x_2} \cdot
n_{y_2}\ket_{\pm} - \bra n_{x_1} \cdot n_{y_1} \ket_{\pm} \bra n_{x_2} \cdot
n_{y_2}\ket_{\pm}$, where $\bra \;\;\ket_{\pm}$ are the functional averages 
with respect to $\cN^{-1}\rme^{-S_{\pm}}$. Note that 
$W_{\pm,r}(\ldots; x,x; \ldots) 
=0$. 

%%%%%%%%%%%%%%%%%%%%%%%%%%%%%%%%%%%%%%%%%%%%%%%%%%%%%%%%%%%%%%%%%%%%%%%
\newsubsection{The $1/N$ expansion}

The goal in the following is to construct these invariant 
correlation functions in a large $N$ asymptotic expansion.   
That is, $\lambda := (N\!+\!1)/\beta$ is kept fixed and the coefficient 
functions $W_{\pm,r}$ in 
\be
W_{\pm,r}(x_1,\dots,y_r) =\frac{\lambda^r}{(N+1)^{r-1}}
\sum_{s=0}^{\infty}\frac{1}{(N+1)^s}W^{(s)}_{\pm,r}(x_1,\dots,y_r)\,,
\label{c5a}
\end{equation}
are sought, with the understanding that the right hand side of 
(\ref{c5a}) provides a valid asymptotic expansion of the exact $W_r$, 
initially on a finite lattice.   

The diagrammatic algorithm for the computation of the coefficient 
functions $W_{+,r}^{(s)}$ is rather straightforward in the compact model, 
see e.g.~\cite{crist}. From \cite{kupi} it is also known to 
provide a valid asymptotic expansion. Direct computation of the functions 
$W_{-,r}^{(s)}$ in the noncompact model is also possible \cite{EMP},
although due to the gauge fixing the 
computations are considerably more tedious than in the compact model. 
In \cite{horo} it will be shown that this algorithm also 
provides a valid asymptotic expansion (\ref{c5a}) for the $W_{-,r}$. 

One of the advantages of a lattice formulation for these 
systems is that there is an {\it exact} correspondence \cite{EMP} 
between the functions $W_{-,r}^{(s)}$ in the noncompact model and 
their counterparts $W_{+,r}^{(s)}$ in the compact model, valid on 
a finite lattice in all dimensions $d \geq 2$:  
  
(a) The coefficient functions $W_{\pm,r}$ are 
translation invariant and can be expressed in terms of  
$D_{\pm}(x) = D(x)|_{\om \ra \om_{\pm}}$, with $D(x)$ the free 
propagator of squared mass $\om$ and with $\om_{\pm}(\lb,V)$ 
the solutions of the gap equations $\lb D(0) = \pm 1$ discussed 
further in Subsection~3.1. 

(b) For all $r \geq 1,\,s \geq 0$, there exists unique 
functionals $X_r^{(s)}[D](\lb)$ of $D$ 
such that $W_{r,+}^{(s)} = X_r^{(s)}[D_+](\lb)$ are the coefficients in 
the compact model and $W_{r,-}^{(s)} = (-1)^r X_r^{(s)}[D_-](-\lb)$
are the coefficients in the noncompact model.   

As a consequence the computations only have to be done in the 
compact model, and the result in the noncompact model can be 
obtained via (b). In the next section we will  
compute a subset of correlation functions, but instead
of doing the computation using the $1/N$ Feynman rules we 
present the results and verify that they solve the 
associated Schwinger-Dyson equations.  
%%%%%%%%%%%%%%%%%%%%%%%%%%%%%%%%%%%%%%%%%%%%%%%%%%%%%%%%%%%%%%%%%%%%
\newpage
\newsection{Schwinger-Dyson Equations} 

In the compact model the Schwinger-Dyson equations for the
functions (\ref{c4}) have, to our knowledge, first been formulated 
by M. L\"{u}scher \cite{luscher}. The derivation is readily extended to 
noncompact models  and is reproduced in \cite{EMP}. Here we just record 
the basic equations: 
\ba 
\label{sd1}
&& \pm \beta \Delta_z \Big[h_{zy} W_\pm - h_{zx} W_\pm - h_{zx} h_{xy} W_\pm 
- (h_{zx} W_\pm)(h_{xy} W_\pm) \Big]\Big|_{z = x} 
\nonum
&& + \sum_{z \neq x} H_{xz} \Big[ h_{zy} W_\pm - h_{zx} W_\pm - h_{xy} W_\pm - 
h_{xz} h_{xy} W_\pm - (h_{zx} W_\pm)(h_{xy} W_\pm) \Big]
\nonum
&& - N (1-\delta_{xy}) (h_{xy} W_\pm +1) =0\,,
\ea 
where the upper sign corresponds to the compact model and the 
lower sign to the noncompact model. 
In terms of the multi-point functions (\ref{c4}) these 
equations amount to an infinite coupled system of nonlinear
partial differential equations. As such boundary conditions have    
to be specified; without them even the exact equations 
(\ref{sd1}) do not determine their solution uniquely, 
see \cite{EMP} for a counter example. However, since 
(\ref{sd1}) does not contain a closed equation for any of the 
$W_r$, it is difficult to impose such boundary conditions in 
practice. 

In contrast, the large $N$ ansatz (\ref{c5a}) effectively
converts the $W_r$ equations into a hierarchy which can be solved
recursively and where `initial' conditions can be specified.
The recursion pattern for the $W_r^{(s)},\, r + s >1$, functions
is given in Fig.~1. To compute a given coefficient all quantities 
having arrows pointing towards it are needed. 

$$
\begin{array}{llll} W_1^{(0)} &\ra W_2^{(0)} &\ra W_3^{(0)} \ra \\[3mm]
\mbox{ } & \;\phantom{\ra} \downarrow & \phantom{\ra} \downarrow \\[3mm]
\mbox{} & \;\phantom{\ra} W_1^{(1)} & \ra W_2^{(1)} \ra \\[3mm] 
\mbox{} & \;\phantom{\ra} \mbox{}  & \phantom{\ra} \downarrow 
\end{array}
$$
\begin{center}
{\small Fig.~1: Recursion pattern for the solution of the large $N$ 
expanded SD equations.} 
\end{center}

The first few equations for the $W_r^{(s)}$ are spelled out in 
Appendix~A; a closed formula for the generic equation can 
also be given and used to show the recursion pattern 
in Fig.~1 by induction. The key assumption in our use 
of the large $N$ expanded Schwinger-Dyson equations will be that at 
each recursion step in Fig.~1 there exists a solution and 
that the solution is {\it unique}. This assumption implies that, 
once $W_1^{(0)}$ has been specified, there will be an infinite 
sequence of functions $W_r^{(s)}$, $r + s >1$, uniquely associated 
with it, which in turn determine the series (\ref{c5a}) uniquely 
for each $W_r$. The choice of $W_1^{(0)}$ is ultimately 
determined by the physics problem one seeks to study; 
different choices are possible \cite{EMP} for the {\it same} 
(initial) equation (\ref{sdk10}).  

The existence and uniqueness of a solution at each recursion step
of Fig.~1 is presumably difficult to establish directly from
the equations. In terms of the underlying discretized 
functional integral (which solves the exact equation (\ref{sd1}) 
by construction) existence and uniqueness of the solution
to the recursive equations amount to the existence of a 
well-defined asymptotic expansion of the form (\ref{c5a}).   
For the generating functionals $W_+[H]$ and $W_-[H]$ the latter 
is guaranteed by the results of \cite{kupi} and \cite{horo},
respectively. This provides an indirect justification 
of the assumption stated in the preceding paragraph. 

It also provides the rationale for the procedure adopted in 
Subsections~3.2 and 3.3: we present the expressions for 
$W_{\pm, 1}^{(s)}$ and $W_{\pm, 2}^{(s)}$ to leading and 
subleading order ($s=0,1$) and claim that they solve 
the Schwinger-Dyson equations with the correct initial conditions,
$W_{1,\pm}^{(0)}$, respectively. The required equations 
are tabulated in Appendix~A. The verification that the 
$W_{\pm, r}^{(s)}$ presented indeed solve these equations
is straightforward and is omitted. It is however far 
shorter than the diagrammatic computation 
in either the compact or the noncompact model. 

Though in this paper we shall be concerned with the large $N$ 
expansion exclusively, let us add that the Schwinger-Dyson equations (\ref{sd1}) can 
also be subjected to a perturbative expansion, i.e.~the 
ansatz (\ref{c5a}) is replaced with one in terms of powers of $1/\beta$. 
The recursive pattern determining the coefficients is similar 
to that in Fig.~1, but the differential part of the equations 
now involves a linear differential operator with constant 
coefficients. This has been used by L\"uscher \cite{luscher} to show 
directly from the equations that each recursion step has at most 
one solution. The existence of a solution of course follows 
from the diagrammatic algorithm as described (on the lattice) 
by Hasenfratz \cite{hasenfratz}. Once the results are known 
the perturbatively expanded Schwinger-Dyson equations provide an efficient way 
to verify them.

%%%%%%%%%%%%%%%%%%%%%%%%%%%%%%%%%%%%%%%%%%%%%%%%%%%%%%%%%%%%%%%%%%%%%%

\newsubsection{2-- and 4-- functions to leading order} 

The leading order two-point functions $W_{\pm, 1}^{(0)}$ provide 
the starting point for the recursion. They have to solve 
Eq.~(\ref{sdk10}) but in order to pick a specific solution 
further information has to be added. The appropriate 
solutions turn out to be given by 
\be
W_{\pm,1}^{(0)}(x,y)=\pm D_\pm(x-y)-\lambda^{-1}\,,
\label{k10}
\end{equation}
with
\be
D_\pm(x)=\frac{1}{V}\sum_p \frac{\rme^{ipx}}{E_p+\om_\pm}\,,
\end{equation}
where the sum goes over 
$p=\frac{2\pi}{L} (n_1,\dots,n_d)\,,\,\,n_\mu=0,1,..,L-1$
and $E_p=\sum_{\mu=1}^d\hat{p}_\mu^2$ with 
$\hat{p}_\mu=2\sin\frac{p_\mu}{2}$. 
Further $\om_\pm$ are the particular solutions to the `gap equations'
\be
\pm\lambda^{-1}=D_\pm(0)=\frac{1}{V}\sum_p\frac{1}{E_p+\om_\pm}\,,
\label{gap}
\end{equation}
obeying $\om_+ >0$ in the compact model and $0> \om_- > - \frac{4}{2 d +1} 
\sin^2 \pi/L$ in the noncompact model \cite{EMP}. The rationale   
for the choice of these solutions of (\ref{sdk10}) is that their 
properties are necessary for the stability of the expansions,
see \cite{kupi} for the compact and \cite{horo} for the 
noncompact model. All $W_{\pm, r}^{(s)}$, $r + s >1$, are then 
in principle uniquely determined by the $W_{\pm, 1}^{(0)}$,
and we shall simply present the solutions of the associated 
Schwinger-Dyson equations.

The solution of (\ref{sdk20}) for the 4--point function in leading 
order is
\ba
&&W_{\pm,2}^{(0)}(x_1,y_1;x_2,y_2)=D_\pm(x_1-x_2)D_\pm(y_1-y_2)
+D_\pm(x_1-y_2)D_\pm(y_1-x_2)
\nonumber\\
&&-2\sum_{u,v}D_\pm(x_1-u)D_\pm(y_1-u)\triangle_\pm(u-v)
D_\pm(v-x_2)D_\pm(v-y_2)\,,
\label{k20}
\ea
where $\triangle_\pm(u-v)$ is defined by 
\be
\sum_u D_\pm(x-u)^2\triangle_\pm(u-v)=\delta_{xv}\,,
\end{equation}
so that
\ba
\triangle_\pm(u) \is \frac{1}{V}\sum_k \frac{\rme^{iku}}{\Pi_\pm(k)}\,,
\nonum
\Pi_\pm(k) &:= &\frac{1}{V}\sum_p \frac{1}{(E_p+\om_\pm)(E_{k+p}+\om_\pm)}\,.
\end{eqnarray} 
Obviously in the compact model $\Pi_+(k)>0$ for all $k$. In contrast 
in the noncompact model one has $\Pi_-(k)<0\,,k\ne0$ and $\Pi_-(0)>0$, 
a fact of importance for the validity of the large $N$ expansion \cite{EMP}.  
$\triangle_\pm(u)$ is the propagator 
of the auxiliary field in the functional treatment of the $1/N$ expansion,
and correspondingly the last term in (\ref{k20}) corresponds to a tree
diagram with an intermediate auxiliary field propagator.

%%%%%%%%%%%%%%%%%%%%%%%%%%%%%%%%%%%%%%%%%%%%%%%%%%%%%%%%%%%%%%%%%%%%%

\newsubsection{2-- and 4--point functions to next-to-leading order} 

The solution of (\ref{sdk11}) for the 2--point function in next-to-leading 
order is 
\ba
W_{\pm,1}^{(1)}(x,y)&=&-2q_\pm\frac{\partial}{\partial\om_\pm}
W_{\pm,1}^{(0)}(x,y)
\nonumber\\
&&\mp 2\sum_{u,v}D_\pm(x-u)D_\pm(u-v)\triangle_\pm(u-v)D_\pm(v-y)\,,
\label{k11}
\ea
where
\ba
\label{qdef}
q_\pm &=& \frac{1}{\Pi_\pm(0)}\sum_{u,v}D_\pm(u)
\triangle_\pm(u-v)D_\pm(u-v)D_\pm(v)
\nonum
&=&\frac{1}{\Pi_\pm(0)V^2}\sum_p\sum_q
\frac{1}{\left(E_p+\om_\pm\right)^2\left(E_{p+q}+\om_\pm\right)\Pi_\pm(q)}\,.
\ea
and the partial derivative $\frac{\partial}{\partial\om_\pm}$
means $\pm\lambda^2\Pi_\pm(0)\frac{\partial}{\partial\lambda}$ (at fixed 
volume). In particular
\ba
\mp\frac{\partial}{\partial\om_\pm}W_{\pm,1}^{(0)}(x,y)
&=&D_{\pm,2}(x-y)
\nonum
&:=&\sum_w D_\pm(w-x)D_\pm(w-y)
=\frac{1}{V}\sum_p\frac{\rme^{ip(x-y)}}{\left(E_p+\om_\pm\right)^2}\,.
\ea
Note $\Pi_\pm(0)=D_{\pm,2}(0)$. The first term on the rhs of 
(\ref{k11}) corresponds to the tadpole diagram and the second 
term to the non-trivial self-energy diagram.

Finally the solution for the next-to-leading 4--point function is
given by
\ba
\nspace &&W_{\pm,2}^{(1)}(x_1,y_1;x_2,y_2)=
-2q_\pm\frac{\partial}{\partial\om_\pm}W_{\pm,2}^{(0)}(x_1,y_1;x_2,y_2)
\nonum
\nspace &&-2\sum_{u,v,w,z}W_{\pm,2}^{(0)}(x_1,y_1;u,v)
D_\pm(u-w)\triangle_\pm(u-w)
D_\pm^{-1}(v-z)W_{\pm,2}^{(0)}(x_2,y_2;w,z)
\nonum
\nspace &&+2\sum_{u,v,w,z}W_{\pm,2}^{(0)}(x_1,y_1;u,v)
D_\pm(u-v)\triangle_\pm(u-w)\triangle_\pm(v-z)
D_\pm(w-z)W_{\pm,2}^{(0)}(x_2,y_2;w,z)
\nonum
\nspace &&-\sum_{u,v}W_{\pm,2}^{(0)}(x_1,y_1;u,v)
\triangle_\pm(u-v)W_{\pm,2}^{(0)}(x_2,y_2;u,v)\,.
\label{k21}
\ea
Again the separate contributions in (\ref{k21}) become more apparent
when drawn as corresponding Feynman diagrams.

%%%%%%%%%%%%%%%%%%%%%%%%%%%%%%%%%%%%%%%%%%%%%%%%%%%%%%%%%%%%%%%%%%%%%

\newsubsection{Two--point current correlation function}

In both models the Noether currents are given by 
\be
J_\mu^{ab}(x)=\beta
\left[n_x^a\partial_\mu n_x^b-n_x^b\partial_\mu n_x^a\right]\,,
\quad 0\leq a,b \leq N\,.
\end{equation}
The invariant two--point function of these currents is 
\be
J_{\pm,\mu\nu}(x,y):=
\left\{ \begin{array}{l} 
\sum_{a,b}\langle 
J_\mu^{ab}(x)J_\nu^{ab}(y)\rangle_+\,,
\\[2mm]
\sum_{a,b,c,d}\langle 
\eta_{ac}\eta_{bd}J_\mu^{ab}(x)J_\nu^{cd}(y)\rangle_-\,.
\end{array}
\right.
\end{equation}
It obeys the Ward identity
\be
\sum_\mu\partial^*_\mu  J_{\pm,\mu\nu}(x,y)=\pm 2N\beta E_\pm
\left(\delta_{x,y}-\delta_{x,y+\hat{\nu}}\right)\,,
\label{WI}
\end{equation}
where $\partial^*_\mu f(x)=f(x)-f(x-\hat{\mu})$ and in the 
noncompact case (\ref{WI}) holds  for $x \neq x_0$ only. 
Further 
\be
E_\pm=\langle n_x\cdot n_{x+\hat{\nu}}\rangle_\pm\,,
\end{equation}
is independent of $x$ because of translation invariance.
One way of obtaining (\ref{WI}) is by specializing the 
`pre-Schwinger-Dyson' equation Eq.~(3.39) of \cite{EMP} to  
${\cal O}=J_\nu^{cd}(y)$ and using the completeness relations  
Eq.~(3.38) in \cite{EMP}.

The two--point function can be expressed in terms of the 
2-- and 4--point functions of the spins according to
\ba
\nspace &&J_{\pm,\mu\nu}(x,y)=2\beta^2
\Bigl[W_{\pm,2}(x,y;x+\hat{\mu},y+\hat{\nu})
                                   -W_{\pm,2}(x,y+\hat{\nu};x+\hat{\mu},y)
\nonum
\nspace &&+W_{\pm,1}(x,y)W_{\pm,1}(x+\hat{\mu},y+\hat{\nu})
  -W_{\pm,1}(x,y+\hat{\nu})W_{\pm,1}(x+\hat{\mu},y)
\nonum
\nspace &&+W_{\pm,1}(x,y)+W_{\pm,1}(x+\hat{\mu},y+\hat{\nu})
-W_{\pm,1}(x,y+\hat{\nu})-W_{\pm,1}(x+\hat{\mu},y)\Bigr]\,.
\ea
The current correlation function has accordingly a $1/N$ expansion of the
form: 
\be
J_{\pm,\mu\nu}(x,y)=2N(N+1)
\sum_{s \geq 0}\frac{1}{(N+1)^s}J_{\pm,\mu\nu}^{(s)}(x,y)\,.
\end{equation}

In the lowest order we have
\ba
\nspace &&J_{\pm,\mu\nu}^{(0)}(x,y)=
\lambda^{-1}\left[W_{\pm,1}^{(0)}(x,y)
+W_{\pm,1}^{(0)}(x+\hat{\mu},y+\hat{\nu})
-W_{\pm,1}^{(0)}(x,y+\hat{\nu})
-W_{\pm,1}^{(0)}(x+\hat{\mu},y)\right]
\nonum
\nspace &&+W_{\pm,1}^{(0)}(x,y)W_{\pm,1}^{(0)}(x+\hat{\mu},y+\hat{\mu})
-W_{\pm,1}^{(0)}(x,y+\hat{\nu})W_{\pm,1}^{(0)}(x+\hat{\mu},y)\,.
\ea
Inserting the solution (\ref{k10}) one gets 
\be
J_{\pm,\mu\nu}^{(0)}(x,y)=
D_\pm(x-y)D_\pm(x+\hat{\mu}-y-\hat{\nu})
            -D_\pm(x-y-\hat{\nu})D_\pm(x+\hat{\mu}-y)\,,
\end{equation}
the Fourier transform of which is 
\ba
\tilde{J}^{(0)}_{\pm,\mu\nu}(q)
&=&\sum_x\rme^{-iqx}J_{\pm,\mu\nu}^{(0)}(x,0)
\nonum
&=&\exp\Big(\frac{i}{2}[q_\mu-q_\nu]\Big)\,\frac{2}{V}\sum_p
\frac{\sin(p+q/2)_\mu\sin(p+q/2)_\nu}
{(E_p+\om_\pm)(E_{p+q}+\om_\pm)}\,.
\label{j0}
\ea
It is seen to satisfy 
\be
\sum_{\mu}\left(1-\rme^{-iq_\mu}\right)
\tilde{J}^{(0)}_{\pm,\mu\nu}(q)
=\left(1-\rme^{-iq_\nu}\right)D_\pm(\hat{\nu})\,,
\end{equation}
which is the Ward identity (\ref{WI}) to lowest order $1/N$.

In next order we have
\ba
\nspace &&J_{\pm,\mu\nu}^{(1)}(x,y)=J_{\pm,\mu\nu}^{(0)}(x,y)
+W_{\pm,2}^{(0)}(x,y;x+\hat{\mu},y+\hat{\nu})
-W_{\pm,2}^{(0)}(x,y+\hat{\nu};x+\hat{\mu},y)
\nonum
\nspace &&+W_{\pm,1}^{(1)}(x,y)\left[W_{\pm,1}^{(0)}(x+\hat{\mu},y+\hat{\nu})
+\lambda^{-1}\right]
  +W_{\pm,1}^{(1)}(x+\hat{\mu},y+\hat{\nu})\left[W_{\pm,1}^{(0)}(x,y)
+\lambda^{-1}\right]
\nonum
\nspace &&-W_{\pm,1}^{(1)}(x,y+\hat{\nu})\left[W_{\pm,1}^{(0)}(x+\hat{\mu},y)
+\lambda^{-1}\right]
  -W_{\pm,1}^{(1)}(x+\hat{\mu},y)\left[W_{\pm,1}^{(0)}(x,y+\hat{\nu})
+\lambda^{-1}\right]\,.
\nonum
\nspace && \mbox{}
\ea
Using the solutions (\ref{k10}), (\ref{k20}), (\ref{k11}), this becomes
\ba
\nspace &&J_{\pm,\mu\nu}^{(1)}(x,y)
=2q_\pm\Bigl[D_{\pm,2}(x-y)D_\pm(x+\hat{\mu}-y-\hat{\nu})
+D_\pm(x-y)D_{\pm,2}(x+\hat{\mu}-y-\hat{\nu})
\nonum
\nspace &&-D_{\pm,2}(x+\hat{\mu}-y)D_\pm(x-y-\hat{\nu})
          -D_\pm(x+\hat{\mu}-y)D_{\pm,2}(x-y-\hat{\nu})\Bigr]
\nonum
\nspace &&-2\sum_w\triangle_\pm(w)D_\pm(w) \Bigl[
D_{\pm,2}(y-x+w) D_\pm(x-y +\hat{\mu}-\hat{\nu})
\nonum
\nspace && + D_{\pm,2}(y-x+\hat{\nu} -\hat{\mu} + w) D_\pm(x-y)
- D_{\pm,2}(y-x-\hat{\mu} +w) D_\pm(x-y -\hat{\nu})
\nonum
\nspace &&
- D_{\pm,2}(y-x+ \hat{\nu} +w) D_\pm(x-y +\hat{\mu}) \Bigr]
%%%%%%%%%%%%%
\nonum
\nspace && -2\sum_{u,v}D_\pm(x-u)D_\pm(y-u)\triangle_\pm(u-v)
D_\pm(x+\hat{\mu}-v)D_\pm(y+\hat{\nu}-v)
\nonum
\nspace &&+2\sum_{u,v}D_\pm(x-u)D_\pm(y+\hat{\nu}-u)\triangle_\pm(u-v)
D_\pm(x+\hat{\mu}-v)D_\pm(y-v)\,.
\ea
Its Fourier transform is simpler in form:
\be 
\tilde{J}^{(1)}_{\pm,\mu\nu}(q)
=-\exp\left(\frac{i}{2}[q_\mu-q_\nu]\right)
\left[X_{\pm,1;\mu\nu}(q)+X_{\pm,2;\mu\nu}(q)+X_{\pm,3;\mu\nu}(q)\right]\,,
\label{j1}
\end{equation} 
with 
\ba
\label{X123def}
X_{\pm,1;\mu\nu}(q)&=&-\frac{8q_\pm}{V}\sum_p 
\frac{\sin(p+q/2)_\mu\sin(p+q/2)_\nu}
{(E_p+\om_\pm)^2(E_{p+q}+\om_\pm)}\,,
\nonum
X_{\pm,2;\mu\nu}(q)&=&\frac{8}{V^2}\sum_{p_1,p_2}
\frac{\sin(p_1+q/2)_\mu\sin(p_1+q/2)_\nu}
{\left(E_{p_1}+\om_\pm\right)^2\left(E_{p_1+q}+\om_\pm\right)
 \left(E_{p_2}+\om_\pm\right)\Pi_\pm(p_1-p_2)}\,, 
\\[2mm]
X_{\pm,3;\mu\nu}(q)&=&\frac{4}{V^2}\sum_{p_1,p_2}
\frac{\sin(p_1+q/2)_\mu\sin(p_2+q/2)_\nu}
{\left(E_{p_1}+\om_\pm\right)\left(E_{p_1+q}+\om_\pm\right)
\left(E_{p_2}+\om_\pm\right)\left(E_{p_2+q}+\om_\pm\right)\Pi_\pm(p_1-p_2)}\,. 
\nonumber
\ea 
It can be checked to satisfy the Ward identity in the next order 
in the large $N$ expansion.

%%%%%%%%%%%%%%%%%%%%%%%%%%%%%%%%%%%%%%%%%%%%%%%%%%%%%%%%%%%%%%%%%%%%%%

\newsubsection{The Binder cumulant}

In scalar field theories with a mass gap a 
renormalized 4--point coupling is defined in terms of 
the Binder cumulant 
\be
\label{gr1} 
U :=  1 + \frac{2}{N\!+\!1} - \frac{\bra (\Sigma \cdot \Sigma)^2 \ket}%
{\bra \Sigma \cdot \Sigma \ket^2}
= - \frac{1}{\bra \Sigma \cdot \Sigma \ket^2} 
\sum_{x_1,x_2,y_1,y_2} \bra n_{x_1}\!\cdot \!n_{y_1} 
n_{x_2}\! \cdot \!n_{y_2} \ket_c\,.
\end{equation} 
Here $\Sigma^a = \sum_x n^a_x$ and 
\ba
\label{gr2}
&& \bra n_{x_1}\cdot n_{y_1} n_{x_2} \cdot n_{y_2} \ket_c
\\
&& \quad = W_2(x_1,y_1; x_2,y_2) 
- \frac{1}{N\!+\!1} 
\bra n_{x_1}\cdot n_{x_2} \ket \bra n_{y_1} \cdot n_{y_2} \ket\,
- \frac{1}{N\!+\!1} 
\bra n_{x_1}\cdot n_{y_2} \ket \bra n_{y_1} \cdot n_{x_2} \ket\,.
\nonumber
\ea
is the usual connected 4--point function, related to the  
previously used second $H$-moment as indicated. In terms of 
$W_1$ and $W_2$ the Binder cumulant reads 
\be 
U_\pm = \frac{2}{N\!+\!1} -  
\dfrac{\sum_{x_1,x_2,y_1,y_2} 
W_{\pm,2}(x_1,y_1;x_2,y_2)}{[\sum_{x,y}(W_{\pm,1}(x,y) +1)]^2}\,.
\label{UW}
\end{equation}   

The Binder cumulant has accordingly a large $N$ expansion of the
form 
\be
\label{gr4}
U_\pm =\sum_{s=0} \frac{1}{(N\!+\!1)^{s+1}} \;U_{\pm,s}(\lb,V) \,.
\end{equation}
The coefficients can obviously be expanded in terms of the coefficients
of the 2-- and 4--point functions summed over all arguments:
\ba
\label{gr3}
w_{\pm,s}(\lb,V)&:=& \sum_{x_1,x_2,y_1,y_2} 
W^{(s)}_{\pm,2}(x_1,y_1;x_2,y_2)\,, 
\nonum
\sigma_{\pm,s}(\lb,V)&:=& 
\sum_{x,y}\left[W^{(s)}_{\pm,1}(x,y)+\lambda^{-1}\delta_{s0}\right]\,.
\ea
The two lowest orders 
\ba 
U_{\pm,0} &= & 2 - \sigma_{\pm,0}^{-2} w_{\pm,0}\,,
\nonum
U_{\pm,1} & = & -\sigma_{\pm,0}^{-2} w_{\pm,1} 
+ 2 \sigma_{\pm,1} \sigma_{\pm,0}^{-3} w_{\pm,0}\,,
\label{u01}
\ea 
involve only functions already computed in the previous subsections. 
We have 
\ba
\label{gr6}
\sigma_{\pm,0}&=& \pm \frac{V}{\om_{\pm}}\,,
\nonum
w_{\pm,0}&=& 2 \sigma_{\pm,0}^2 - \frac{2V}{\om_{\pm}^4\Pi_\pm(0)}\,,
\ea
and
\ba
\label{gr10}
\sigma_{\pm,1}&=& \pm \frac{2V}{\om_{\pm}^2} 
[ q_{\pm} - \sum_x \Delta_{\pm}(x) D_{\pm}(x)]\,,
\nonum
w_{\pm,1} &=& -2 q_\pm \frac{\dd}{\dd \om_\pm} w_{\pm,0} + 
2\sum_{u,v,z,w}r_\pm(u,v)r_\pm(w,z) 
\\
&& \times \Delta_\pm(u-w) [-D_\pm(u-w) D_\pm^{-1}(v-z) 
+ D_\pm(u-v) \Delta_\pm(v-z) D_\pm(w-z)]
\nonum
&& 
-\sum_{u,v} r_\pm(u,v)^2 \Delta_\pm(u-v)\,,
\nonumber
\ea
where
\be
\label{gr9} 
r_\pm(x,y):=\sum_{z,w} W_{\pm,2}^{(0)}(z,w;x,y) = 
\frac{2}{\om^2_{\pm}}\Big[1 - \Pi_\pm(0)^{-1} D_{\pm,2}(x-y)\Big]\,.
\end{equation}

%%%%%%%%%%%%%%%%%%%%%%%%%%%%%%%%%%%%%%%%%%%%%%%%%%%%%%%%%%%%%%%%%%%%%%%%%%%

\newpage
\newsection{TD limit of spin and current two-point functions}

Up to this point we have been considering both the compact and noncompact
models in arbitrary dimensions $d \geq 2$. In the following we restrict 
attention to $d=2$. Also numerous articles have dealt with the $1/N$ expansion  
of the compact model so we will in this and in the next section restrict 
attention to the noncompact case and drop the minus $(-)$ suffix on all 
functions.

The results summarized in Section~2 seemingly suggest a simple relation 
between the compact and the noncompact models. It is important to 
stress, however, that the relations hold only on a finite lattice and 
for invariant correlators. Physical quantities arising after 
taking the limit of infinite lattice size (thermodynamic limit)  
turn out to be very different in both systems. We will illustrate this 
fact by studying the thermodynamic (TD) limit of the coefficients in 
the $1/N$ expansions of the correlators computed in the last section 
in the noncompact model. The very existence of the limit is non-trivial
in this case because $\om\to 0$ as $V\to\infty$,
specifically 
\be 
\om = \om_-(\lb, V) \sim - \frac{4\pi}{V \ln V} + 
\rmO\Big(\frac{1}{V \ln^2 V}\Big)\,.   
\end{equation}
This means a `coordinated' limit of lattice sums of the form
$\frac{1}{V^n} \sum_{k_1,\ldots, k_n} f_V(k_1, \ldots, k_n)$ 
has to be taken, where $f_V$ via $\om$ carries an explicit $V$-dependence.
The gap equation (\ref{gap}) effectively acts as a subtle 
infrared regulator whose usefulness is underlined by the 
result summarized in Subsection~2.2. As mentioned, the sums 
associated with individual Feynman diagrams will typically diverge 
in the limit. The issue is whether the infrared divergences cancel 
in the $W_r^{(s)}$ and the quantities computed in terms of them.   

In this section we discuss the limit of the spin and current 
two--point functions; the limit of the Binder cumulant 
is computed in Section~5.

%%%%%%%%%%%%%%%%%%%%%%%%%%%%%%%%%%%%%%%%%%%%%%%%%%%%%%%%%%%%%%%%%%%%%%%%%%%%

\newsubsection{TD limit of the spin two--point function}

In the leading order the 2--point function has an infinite volume limit
\be
-W_1^{(0)}(x,0)=\frac{1}{V}\sum_{p\ne0}\frac{\rme^{ipx}-1}{E_p+\om}
\;\;\rra \;\; {\cal D}(x):=\int_p\,\frac{\rme^{ipx}-1}{E_p}\,,
\end{equation}
where here and in the following $\int_p$ means integration over the
Brillouin zone $\int_0^{2\pi}\frac{\rmd^2p}{(2\pi)^2}$.
The infinite volume lattice propagator ${\cal D}(x)$ is a remarkable
function which has been discussed in detail by Shin \cite{shin}.
At every lattice point it is given by an expression of the form
$r_1(x)+r_2(x)/\pi$ where $r_i$ are rational numbers. 
As $|x|\to\infty$ it diverges logarithmically:
\be
{\cal D}(x)\sim -\frac{1}{4\pi}\left(\ln x^2+2\gamma+3\ln2\right)
+{\rm O}(|x|^{-2})\,,
\end{equation}  
where $\gamma \approx 0.577$ is Euler's constant. 

The next-to-leading term is given by 
\be 
W_1^{(1)}(x,0)= \frac{1}{V} \sum_{p \neq 0} (\rme^{ipx}-1)W_1^{(1)}(p)\,,
\end{equation}
with
\be
W_1^{(1)}(p)  = \frac{2}{(E_p + \om)^2} \Big[ 
- q + \frac{1}{V} \sum_k \frac{1}{\Pi(k) (E_{p-k} + \om)} \Big]\,,
\label{two2}
\end{equation} 
and $q=q_-$ as in (\ref{qdef}). 
For $p=0$ the TD limit of $W_1^{(1)}(p)$ does not exist (similarly to the
situation for the leading order): 
$W_1^{(1)}(0)\sim V \ln V \ln \ln V$, 
reflecting the fact that $W_1(x,0)$ is an increasing 
function of distance $|x|$.

For $p \neq 0$, however the limit exists. To see this we first note 
that $\Pi(p)$ has a TD limit for $p\ne0$. Indeed
using one insertion of (\ref{binder7}b) below and the gap equation
one can rewrite $\Pi(p)$ as 
\ba 
\Pi(p) \is -\frac{1}{(E_p + 2\om)}\Big[\frac{2}{\lb} + J(p) \Big]\,,
\quad p \neq 0\,,
\nonum
J(p) &:= & \frac{1}{V} \sum_k 
\frac{ E_k + E_{p-k}-E_p }{(E_k + \om)(E_{p-k} +\om)}\,.
\label{two6}
\ea   
Throughout we often use the symbol $J$ to denote lattice sums 
which give rise to convergent integrals over the Brillouin zone 
upon taking the infinite volume limit. The limit $\Pi_\infty(p)$ of $\Pi(p)$
is then given by
\be
\Pi_\infty(p) = -\frac{1}{E_p\,v(p)}\,,\quad p \neq 0\,,
\sspace 
v(p) := \left[\frac{2}{\lb} + J_\infty(p)\right]^{-1}\,,
\label{piinf}
\end{equation}
with
\be 
J_{\infty}(p) = \int_k \frac{ E_k + E_{p-k}- E_p  }{E_k E_{p-k}}\,.
\label{two7}
\end{equation}

The properties of the function $J_{\infty}(p)$ will be important 
later on; we mostly need: 
\be 
J_{\infty}(p) \geq 0\,\,\,\,{\rm for}\,\,p\ne0\,, 
\quad J_{\infty}((\pi,\pi))=0\,,\quad
\label{two8a}
\end{equation}
and the behavior for $p\to0$ 
\be 
J_{\infty}(p) = -\frac{1}{2\pi} [\ln p^2 - 5 \ln 2] 
+ {\rm O}(p^2 \ln p^2)\,.    
\label{two8}
\end{equation}
Eq.~(\ref{two8}) follows from \cite{shin}, the positivity 
in (\ref{two8a}) follows from Appendix~A of \cite{EMP}. 
A direct way to see $J_{\infty}(p) \geq 0$ is by performing 
one of the integrations explicitly. This leads to 
the integral representation
\ba
J_\infty(p)=\frac12\int_0^{2\pi}\frac{\rmd k_1}{2\pi}
\left[- {\rm th} \frac{a}{2} - {\rm th} \frac{b}{2} 
+ \frac{\sh(a+b)}{\sh a\,\sh b}\left(
\frac{4\sh^2\left(\frac{a+b}{2}\right)-\hat{p}_1^2}
     {4\sh^2\left(\frac{a+b}{2}\right)+\hat{p}_2^2}\right)\right]\,,
\label{jinf}
\ea
where $a,b>0$ are determined by $\sh \frac{a}{2} = |\sin \frac{k_1}{2}|$,
$\sh \frac{b}{2} = |\sin \frac{k_1-p_1}{2}|$. For the numerator 
of the last term in the integrand one then has 
\ba
&&4\sinh^2\left(\frac{a+b}{2}\right)-\hat{p}_1^2
\nonumber\\
&&=8\left[\sinh\frac{a}{2}\sinh\frac{b}{2}\cosh\left(\frac{a+b}{2}\right)
+\sin\left(\frac{k_1}{2}\right)\sin\left(\frac{k_1-p_1}{2}\right)
\cos\left(\frac{p_1}{2}\right)\right]
\nonumber\\
&&=8\sinh\frac{a}{2}\sinh\frac{b}{2}\left[\cosh\left(\frac{a+b}{2}\right)
+\epsilon\cos\left(\frac{p_1}{2}\right)\right]\,,
\ea
where $\epsilon=\pm1$. For fixed $p_1,k_1$ the integrand of 
(\ref{jinf}) therefore is a monotonically decreasing function of $p_2$ for 
$0<p_2<\pi$. By symmetry the same must hold with the roles 
of $p_1$ and $p_2$ interchanged, so that $J_\infty(p)\ge 
J_\infty((\pi,\pi))=0$, for all $p \neq 0$.

Returning to (\ref{two2}) we rewrite the sum as 
\ba 
\label{two3}
&& \frac{1}{V} \sum_k \frac{1}{\Pi(k) (E_{p-k} + \om)}  
=  J_\Pi(p) - \frac{1}{\lb\Pi(p)}\,,
\\ 
&&  J_\Pi(p) := 
\frac{1}{2 V} \sum_k \frac{1}{E_k + \om} \Big( \frac{1}{\Pi(p-k)} 
+   \frac{1}{\Pi(p+k)}  - \frac{2}{\Pi(p)} \Big)\,. 
\nonumber
\ea
Then using the properties of the $\Pi$ function (\ref{two8a}) and 
(\ref{two8}), it follows that $J_\Pi$ has a finite TD limit which we 
denote by $J_{\Pi,\infty}$. In Section~5 we show that also the TD
limit of $q$ exists,
\be
q_\infty=-\int_k v(k)=J_{\Pi,\infty}(0)\,.
\label{two4}
\end{equation}
Putting the results together one sees that the limit of (\ref{two2}) is
\ba
\label{two5}
W_{1,\infty}^{(1)}(p) \is 
\frac{2}{E_p}\left[j(p)+\lambda^{-1}v(p)\right]\,, \quad p \neq 0\,,
\\
j(p) &:= &\frac{1}{E_p}\left[J_{\Pi,\infty}(p)-J_{\Pi,\infty}(0)\right]\,.
\nonumber
\ea 
From (\ref{two3}) one also gets the `sum rule' 
\be 
\int_p E_p W_{1,\infty}^{(1)}(p) = 2 + \frac{2}{\lb} q_{\infty}\,,
\end{equation}
while $\int_p W_{1,\infty}^{(1)}(p) = 0 = W_{1,\infty}^{(1)}(x,x)$, as required. 

It is instructive to compare now the continuum (i.e.~small $ap$ behavior
where $a$ is the lattice spacing) of $W_{1,\infty}^{(1)}(p)$ with 
its counterpart in the compact model. In Appendix~B the small $p$ asymptotics 
of $j(p)$ is determined. In terms of the (non-universal) constants 
$g_2(\lb),\,g_3(\lb)$ the small $p$ behavior comes out as 
\be
E_p W^{(1)}_{1,\infty}(p) \sim -\ln\left(-\ln(p^2/T)\right)+ 2g_2(\lb)
+\frac{2 g_3(\lb) - 4\pi/\lb}{\ln(p^2/T)}+\rmO\left([\ln(p^2/T)]^{-2}\right)\,,
\label{smallplatt}
\end{equation}
where 
\be
T=32\exp\left(\frac{4\pi}{\lambda}\right)\,.
\label{defT}
\end{equation}
Eqs.~(\ref{smallplatt}), (\ref{defT}) illustrate in particular the 
nonperturbative nature of 
the large $N$ expansion in the noncompact model, despite the fact 
that in infinite volume the expansion is effectively performed with 
respect to massless fields, as it is the case in perturbation theory. 
The infrared regulator $\om(\lb,V)$ clearly works very differently 
from a constant `small mass' regulator. 

The subleading logarithmic terms in (\ref{smallplatt}) are difficult to 
determine on the lattice. Using a continuum cutoff instead and 
cutoff-normalized continuum momenta  
\be 
\frac{1}{V} \sum_k \mapsto \int \!\frac{\rmd^2 k}{(2\pi)^2}\, 
\theta(\Lambda^2 - k^2) \,,
\sspace [-\pi, \pi] \ni p_{\rm latt} \mapsto  
p_c := \frac{\sqrt{32} p}{\Lambda}\,,
\end{equation}
the continuum counterpart of (\ref{two5}) can be found in 
closed form: 
\ba
E_{p_c} W_{1,c}(p_c) \is 
- \frac{2 T}{p_c^2} \,{\rm Li}\Big(\frac{p_c^2}{T}\Big)
- 1 + \frac{4\pi}{\lambda} \frac{1}{\ln(T/p_c^2)}
\nonum
&& -\frac{1}{\ln (T/p_c^2)}
\int_{0}^{32/p_c^2} \rmd s\,\frac{\ln s}{|1-s|}
\frac{1}{\ln(T/p_c^2) -\ln s}\,, 
\label{smallpcont1}
\ea 
where ${\rm Li}(x) = \int_0^x \rmd s/\ln s$. The  
integral in the last term is singularity free as the divergence of
the $1/|1-s|$ factor at $s=1$ is removed by the $\ln s$, 
and $\ln T/p_c^2 > \ln s$ holds over the entire range of the integration.
The asymptotic expansion of (\ref{smallpcont1}) comes out as 
\ba 
\label{smallpcont2}
E_{p_c} W_{1,c}(p_c) & \sim & 
- \ln\Big(\! -\!\frac{\lb}{4\pi} \ln( p_c^2/T) \Big) + 
\frac{2}{\ln (T/p_c^2)}  
\nonum
&& + \sum_{n \geq 0} c_{n}\frac{1}{\ln^{n+2}(T/p_c^2)} + {\rm O}(p_c^2)\,,
\nonum
c_{n} &:=& (-1)^{n+1}\Gamma(n+2)[2 - ((-1)^{n+1} +1) \zeta(n+2)]\,. 
\ea 
We add some remarks. First, note that (\ref{smallpcont2}) 
contains factorially growing terms both with oscillating signs 
(Borel-summable) and with constant phase (non-Borel-summable). 
Second, (\ref{smallpcont2}) is an expansion in terms of the 
running coupling 
\be 
\frac{\alpha(p)}{2\pi} := 
\frac{1}{ \ln (T/p_c^2)} = \frac{\frac{\lb}{4\pi}}{1 + \frac{\lb}{4\pi} \ln 
\frac{\Lambda^2}{p^2}}\,,
\end{equation} 
which can be re-expanded in terms of positive powers of 
the bare coupling $\lambda$. Comparing the result with its counterpart 
in the compact model one may verify that both perturbative expansions are 
related simply by flipping the sign of $\lb$; see \cite{EMP} 
for a general proof of this ``perturbative correspondence''. 
Third, the expressions (\ref{smallplatt}), (\ref{smallpcont2}) 
do not suggest a nontrivial continuum limit reachable merely 
by a multiplicative field and a coupling renormalization. 
For example defining a renormalized coupling 
$\lb_r$ by $\Lambda^2 \rme^{4\pi/\lb} = \mu^2 \rme^{4\pi/\lb_r}$, 
the infinite cutoff limit can only be taken by allowing 
negative bare couplings. For all $\lb >0$ both $\lb_r$ and 
$\alpha(p)$ vanish for $\Lambda \ra \infty$.   
Finally, the expressions  (\ref{smallpcont1}), (\ref{smallpcont2}) can 
also directly be compared with their analogues in the compact model, 
see \cite{Flyvbjerg} for the former. In the 
compact model the bare mass gap $m_0^2 = \Lambda^2 \rme^{-4\pi/\lb}$ enters
and for $m_0^2/p^2 =  32 \rme^{-4\pi/\lb}/p_c^2 \ll 1$ (i.e.~small 
$\lb$ at fixed $p_c$) expressions for the self energy 
$E_{p_c} W_{1,c}(p_c)$ are obtained related to  (\ref{smallpcont1}),
(\ref{smallpcont2}) formally by flipping the sign of $\lb$. 
The attempt to take the sign flip beyond the asymptotic 
small $\lb$ expansion, however, would on the bare level produce a 
``hyperviolet'' mass scale $\Lambda^2 \rme^{+4\pi/\lb}$, 
difficult to interpret. In contrast, in a lattice formulation the 
exact ``large $N$ correspondence'' summarized in Subsection 2.2 exists.

%%%%%%%%%%%%%%%%%%%%%%%%%%%%%%%%%%%%%%%%%%%%%%%%%%%%%%%%%%%%%%%%%%%%%%%%%%%%%%%

\newsubsection{TD limit of the Noether current two--point function}

To obtain the infinite volume limit of the leading order contribution to the
current correlation function (\ref{j0}) we decompose it according to:
\be
\tilde{J}^{(0)}_{\mu\nu}(q)
=2\exp\left(\frac{i}{2}[q_\mu-q_\nu]\right)
\left[A_{\mu\nu}(q)+\sin(q_\mu/2)\sin(q_\nu/2)\Pi(q)\right]\,,
\end{equation}
with
\be
A_{\mu\nu}(q)=\frac{1}{V}\sum_p
\frac{\sin(p+q/2)_\mu\sin(p+q/2)_\nu-\sin(q_\mu/2)\sin(q_\nu/2)}
{(E_p+\omega)(E_{p+q}+\omega)}\,.
\end{equation}
Now we have seen in the last subsection that $\Pi(q)$, $q\ne0$, has a 
finite limit, and so obviously does $A_{\mu\nu}(q)$:
\be
A_{\mu\nu}(q)\to A_{\infty,\mu\nu}(q)=\int_p\,\frac{1}{E_pE_{p+q}}
\left[\sin(p+q/2)_\mu\sin(p+q/2)_\nu-\sin(q_\mu/2)\sin(q_\nu/2)\right]\,,
\end{equation}
which is independent of $\lambda$. Thus the infinite volume limit of 
$J_{\mu\nu}^{(0)}(q)$ exists.

In the next order we have the result (\ref{j1}). We now 
show consecutively: (i) that $X_3$ has a finite TD limit, and (ii) 
that $X_1+X_2$ has a finite TD limit.

(i) Noting the identity
\be
\sum_{p}\frac{\sin(p+q/2)_\mu}
{\left(E_{p}+\omega\right)\left(E_{p+q}+\omega\right)}=0\,,
\end{equation}
we can write $X_3$ in (\ref{X123def}) as
\be
X_{3;\mu\nu}(q)=\frac{4}{V^2}\sum_{p_1,p_2}
\frac{\sin(p_1+q/2)_\mu\sin(p_2+q/2)_\nu Y(p_1,p_2)}
{\left(E_{p_1}+\omega\right)\left(E_{p_1+q}+\omega\right)
 \left(E_{p_2}+\omega\right)\left(E_{p_2+q}+\omega\right)}\,, 
\end{equation}
where
\be
Y(p_1,p_2)=\frac{1}{\Pi(p_1-p_2)}-\frac{1}{\Pi(p_1)}-\frac{1}{\Pi(p_2)}\,,
\end{equation}
which vanishes when either $p_1=0$ or $p_2=0$. Then we break up 
$X_3$
\be
X_{3;\mu\nu}=\frac{1}{(E_q+\omega)^2}
\left[X_{4;\mu\nu}-X_{5;\mu\nu}-X_{5;\nu\mu}-X_{6;\mu\nu}\right]\,,
\end{equation}
with
\ba
\nspace X_{4;\mu\nu}(q) &=& \frac{4}{V^2}\sum_{p_1,p_2}
\frac{\sin(p_1+q/2)_\mu\sin(p_2+q/2)_\nu Y(p_1,p_2)
(E_q-E_{p_1})(E_q-E_{p_2})}
{\left(E_{p_1}+\omega\right)\left(E_{p_1+q}+\omega\right)
 \left(E_{p_2}+\omega\right)\left(E_{p_2+q}+\omega\right)}\,, 
\nonum
\nspace X_{5;\mu\nu}(q) &=& \frac{4}{V^2}\sum_{p_1,p_2}
\frac{\sin(p_1+q/2)_\mu\sin(p_2+q/2)_\nu Y(p_1,p_2)
(E_q-E_{p_1})}
{\left(E_{p_1}+\omega\right)\left(E_{p_1+q}+\omega\right)
 \left(E_{p_2+q}+\omega\right)} \,,
\\[2mm]
\nspace X_{6;\mu\nu}(q) &=& \frac{4}{V^2}\sum_{p_1,p_2}
\frac{\sin(p_1+q/2)_\mu\sin(p_2+q/2)_\nu Y(p_1,p_2)}
{\left(E_{p_1+q}+\omega\right)\left(E_{p_2+q}+\omega\right)}\,. 
\nonumber
\ea 
Now the limit of $X_4$ exists. Next for $X_5$ we write
\be
X_{5;\mu\nu}(q)=X_{7;\mu\nu}(q)+X_{8;\mu\nu}(q)\,,
\end{equation}
with
\ba
X_{7;\mu\nu}(q)&=&
\frac{4}{V^2}\sum_{p_1,p_2}
\frac{\sin(p_1+q/2)_\mu\sin(p_2+q/2)_\nu\left[Y(p_1,p_2)-Y(p_1,-q)\right]
(E_q-E_{p_1})}
{\left(E_{p_1}+\omega\right)\left(E_{p_1+q}+\omega\right)
\left(E_{p_2+q}+\omega\right)}\,,
\nonum
X_{8;\mu\nu}(q)&=&r_\nu(q)\frac{1}{V}
\sum_{p}\frac{\sin(p+q/2)_\mu Y(p,-q)(E_q-E_p)}
{\left(E_{p}+\omega\right)\left(E_{p+q}+\omega\right)}\,,
\ea
where (using the gap equation)
\be
r_\mu(q):=\frac{1}{V}\sum_{p}
\frac{\sin(p+q/2)_\mu}{E_{p+q}+\omega }
=\frac14\left[1+(4+\omega)\lambda^{-1}\right]\sin(q_\mu/2)\,.
\end{equation}
It is clear that both $X_7, X_8$ and so also $X_5$ have finite limits. 
Finally we have
\be
X_{6;\mu\nu}(q)=X_{9;\mu\nu}(q)+X_{10;\mu\nu}(q)+X_{10;\nu\mu}(q)
+4Y(q,q)r_\mu(q)r_\nu(q)\,,
\end{equation}
with
\ba
X_{9;\mu\nu}(q)&=&
\frac{4}{V^2}\sum_{p_1,p_2}\sin(p_1+q/2)_\mu\sin(p_2+q/2)_\nu
\nonum
&&\times\frac{[Y(p_1,p_2)-Y(p_1,-q)-Y(p_2,-q)+Y(q,q)]}
{\left(E_{p_1+q}+\omega\right)\left(E_{p_2+q}+\omega\right)} 
\\[2mm]
X_{10;\mu\nu}(q)&=&r_\mu(q)\frac{4}{V}\sum_{p}
\frac{\sin(p+q/2)_\nu [Y(p,-q)-Y(q,q)]}{E_{p+q}+\omega}\,. 
\nonumber
\ea 
It follows that $X_8$ has a limit, and the demonstration that $X_3$
has a finite TD limit is complete.

(ii) We first rewrite the sum
\be
X_{1;\mu\nu}+X_{2;\mu\nu}=-8q_-A_{2;\mu\nu}+X_{11;\mu\nu}\,,
\end{equation}
with
\ba
A_{2;\mu\nu}(q)&=&\frac{1}{V}\sum_p \frac{F_{\mu\nu}(p,q)}{(E_p+\omega)^2}\,,
\nonum
X_{11;\mu\nu}(q)&=&\frac{8}{V^2}\sum_{p_1,p_2}
\frac{F_{\mu\nu}(p_1,q)}
{\Pi(p_1-p_2)\left(E_{p_1}+\omega\right)^2\left(E_{p_2}+\omega\right)}\,.
\ea
Here 
\be
F_{\mu\nu}(p,q)=\frac{\sin(p+q/2)_\mu\sin(p+q/2)_\nu}{E_{p+q}+\omega}
-\frac{\sin(q/2)_\mu\sin(q/2)_\nu}{E_q+\omega}\,.
\end{equation}
To proceed we again write $X_{11}$ as a sum of terms:
\be
X_{11;\mu\nu}(q)=X_{12;\mu\nu}(q)+X_{13;\mu\nu}(q)+8f_1 A_{2;\mu\nu}(q)\,,
\end{equation}
where 
\ba
X_{12;\mu\nu}(q)&=&\frac{8}{V^2}\sum_{p_1,p_2}
\frac{Y(p_1,p_2)F_{\mu\nu}(p_1,q)}
{\left(E_{p_1}+\omega\right)^2\left(E_{p_2}+\omega\right)}\,,
\nonum
X_{13;\mu\nu}(q)&=&-\frac{8\lambda^{-1}}{V}\sum_{p}
\frac{F_{\mu\nu}(p,q)}{\Pi(p)\left(E_{p}+\omega\right)^2}\,,
\ea
and ($f_2$ will appear in the next section):
\be
f_s:=\frac{1}{V}\sum_{p}
\frac{1}{\Pi(p)\left(E_{p}+\omega\right)^s}\,.
\label{fs}
\end{equation}
We still need to consider the limits of
$X_{12;\mu\nu},X_{13;\mu\nu},A_{2;\mu\nu}$ (actually the contribution in 
$X_1+X_2$ involving $A_2$ has a coefficient proportional to $q-f_1$ which 
vanishes in the TD limit). 
We note that these three functions can be written in the form
$T_{2;\mu\nu}[g](q)$ where functions $T_{s;\mu\nu}[g](q)$
are defined by
\be
T_{s;\mu\nu}[g](q)=\frac{1}{V}\sum_p 
\frac{g(p)F_{\mu\nu}(p,q)}{(E_p+\omega)^s}\,,
\end{equation}
with $g(p)$ a regular periodic function with $g(p)=g(-p)$ and 
finite at $p=0$.
Now for functions of the type $T_1$ we have
\ba
&&T_{1;\mu\nu}[g](q)=\frac{1}{V}\sum_p 
\frac{g(p)F_{\mu\nu}(p,q)}{E_p+\omega}
\nonum
&&= \frac{1}{E_q+2\omega}
\Bigl\{\frac{1}{V}\sum_p\frac{g(p)F_{\mu\nu}(p,q)(E_q-E_p-E_{p+q})}
{E_p+\omega}
\nonumber\\
&&\;\;+\frac{1}{V}\sum_p\frac{g(p)F_{\mu\nu}(p,q)(E_{p+q}+\omega)}{E_p+\omega}
\nonumber\\
&&\;\;+\frac{1}{V}\sum_p [g(p)-g(q)]F_{\mu\nu}(p,q)
+g(q)\frac{1}{V}\sum_p F_{\mu\nu}(p,q)\Bigr\}\,,
\ea
which has a finite limit since $\frac{1}{V}\sum_p F_{\mu\nu}(p,q)$ 
does, as can easily be seen using the gap equation. Next
\ba
&&T_{2;\mu\nu}[g](q)=\frac{1}{V}\sum_p 
\frac{g(p)F_{\mu\nu}(p,q)}{(E_p+\omega)^2}
\nonum
&& = \frac{1}{E_q+2\omega}
\left\{T_{1;\mu\nu}[g](q)+T_{3;\mu\nu}[g](q)+T_{4;\mu\nu}[g](q)\right\}\,.
\ea
Here 
\ba
&& T_{3;\mu\nu}[g](q)=
\frac{1}{V}\sum_p\frac{g(p)F_{\mu\nu}(p,q)(E_q-E_p-E_{p+q})}
{(E_p+\omega)^2}\,,
\nonum
&& T_{4;\mu\nu}[g](q)=\frac{1}{(E_q+\omega)}
\frac{1}{V}\sum_p\frac{g(p)\overline{F}_{\mu\nu}(p,q)}{(E_p+\omega)^2}\,,
\ea
with
\be
\overline{F}_{\mu\nu}(p,q) := 
(E_q+\omega)(E_{p+q}+\omega)F_{\mu\nu}(p,q)\,.
\end{equation}
Decomposing $T_3$ further gives 
\be
T_{3;\mu\nu}[g](q)=\frac{1}{E_q+\omega}\left\{
T_{5;\mu\nu}[g](q)+T_{6;\mu\nu}[g](q)\right\}\,,
\end{equation}
with
\ba
&&T_{5;\mu\nu}[g](q)=
\frac{1}{V}\sum_p\frac{g(p)F_{\mu\nu}(p,q)(E_q-E_p-E_{p+q})(E_q-E_{p+q})}
{(E_p+\omega)^2}
\nonum
&&T_{6;\mu\nu}[g](q)=\frac{1}{E_q+\omega}
\frac{1}{V}\sum_p\frac{g(p)\bar{F}_{\mu\nu}(p,q)(E_q-E_p-E_{p+q})}
{(E_p+\omega)^2}\,,
\ea
where $T_5$ clearly has a TD limit. Now 
\be
\bar{F}_{\mu\nu}(p,q)=\bar{F}_{-;\mu\nu}(p,q)+\bar{F}_{+;\mu\nu}(p,q)\,,
\end{equation}
with $F_\pm(-p,q)=\pm F_\pm(p,q)$:
\begin{subeqnarray}
\nspace && \bar{F}_{-;\mu\nu}(p,q)
=(E_q+\omega)\left[\sin p_\mu\cos p_\nu\cos(q/2)_\mu\sin(q/2)_\nu
+(\mu\leftrightarrow\nu)\right]
\nonum
\nspace && \quad -2\sum_\rho\sin p_\rho\cos q_\rho\sin(q/2)_\mu\sin(q/2)_\nu\,,
\\
\nspace &&\bar{F}_{+;\mu\nu}(p,q)
=(E_q+\omega)\sin p_\mu\sin p_\nu\cos(q/2)_\mu\cos(q/2)_\nu
\nonum
\nspace &&\quad -\frac12\sin(q/2)_\mu\sin(q/2)_\nu\Big[
(E_q+\omega)(\hat{p}_\mu^2+\hat{p}_\nu^2-\frac12\hat{p}_\mu^2\hat{p}_\nu^2)
+2E_p-\sum_\rho\hat{p}_\rho^2\hat{q}_\rho^2\Big]\,.
\end{subeqnarray} 
So
\be
T_{6;\mu\nu}[g](q)=
\frac{1}{E_q+\omega}\left\{T_{7;\mu\nu}[g](q)+T_{8;\mu\nu}[g](q)
+g(0)T_{9;\mu\nu}(q)\right\}\,,
\end{equation}
with
\ba
&&T_{7;\mu\nu}[g](q)=\frac12\sum_\rho\hat{q}_\rho^2
\frac{1}{V}\sum_p\frac{g(p)\bar{F}_{+;\mu\nu}(p,q)\hat{p}_\rho^2}
{(E_p+\omega)^2}\,,
\nonum
&&T_{8;\mu\nu}[g](q)=-2\sum_\rho\cos q_\rho
\frac{1}{V}\sum_p\frac{[g(p)-g(0)]\bar{F}_{-;\mu\nu}(p,q)\sin p_\rho}
{(E_p+\omega)^2}\,,
\\[2mm]
&&T_{9;\mu\nu}(q)=-2\sum_\rho\cos q_\rho
\frac{1}{V}\sum_p\frac{\bar{F}_{-;\mu\nu}(p,q)\sin p_\rho}
{(E_p+\omega)^2}\,.
\nonumber
\ea
All of these quantities have a finite limit, in particular $T_{9;\mu\nu}(q)$
on account of the gap equation. A similar decomposition can be performed 
for $T_{4;\mu\nu}(q)$, showing that it likewise has a TD limit. 
So $T_{2;\mu\nu}[g](q)$ has a TD limit. It follows that 
$X_{12;\mu\nu},X_{13;\mu\nu},A_{2;\mu\nu}$ all have infinite volume limits.

To summarize, we have shown that the TD limit of the contribution to the 
two leading orders in the $1/N$ expansion of the current correlator 
$J^{(s)}_{\mu\nu}\,,s=0,1$, exists.

%%%%%%%%%%%%%%%%%%%%%%%%%%%%%%%%%%%%%%%%%%%%%%%%%%%%%%%%%%%%%%%%%%%%%%%%%%%%%%%

\newsection{TD limit of the Binder cumulant}

In the following we evaluate the infinite volume limit of $U$ in the 
2-dimensional noncompact model to sub-leading order, i.e.~the 
coefficients $U_0$, $U_1$ in 
\be 
U(\lb,V) = \frac{U_0(\lb,V)}{N\!+\!1} + \frac{U_1(\lb,V)}{(N\!+\!1)^2} 
+ {\rm O}\Big(\frac{1}{(N\!+\!1)^3}\Big)\,, 
\end{equation}
are computed for large volumes. Somewhat surprisingly we   
will find that the limit $V \ra \infty$ exists and is 
independent of $\lb$! In Subsection~5.1 the large volume asymptotics 
will be evaluated analytically. As a test of the estimates 
and in order to have finite volume results to compare Monte-Carlo 
data with, we also directly evaluated the multiple 
lattice sums numerically up to $L=1024$. The results are 
reported in Subsection~5.2. Finally, to preclude that the 
large $N$ results are misleading we performed a Monte-Carlo 
study of $U(\lb,V)$ up to $L=384$ for $N=8$.

%%%%%%%%%%%%%%%%%%%%%%%%%%%%%%%%%%%%%%%%%%%%%%%%%%%%%%%%%%%%%%%%
\newsubsection{Analytical analysis of large $V$ asymptotics} 

Evaluation of the leading order coefficient is straightforward. 
From (\ref{u01}) and (\ref{gr6}) one obtains 
\be
\label{gr7}
U_0(\lb,V) = \frac{2}{V \om^2\Pi(0)}\,.
\end{equation} 
Now the TD limit of $\om \Pi(0)$ can be evaluated from 
\be 
- \om \Pi(0) = \frac{1}{\lb} + \frac{1}{V} \sum_{k \neq 0} 
\frac{E_k}{(E_k + \om)^2} = 
\frac{1}{4\pi} \ln V + \frac{1}{\lb} + a(\lb) + 
\rmO\Big(\frac{1}{\ln V}\Big) \,,
\label{pi01}
\end{equation}
where $a(\lb) > 0$. Note that (taking the $\lb$-derivative of the 
gap equation (\ref{gap})) this translates into a large volume 
asymptotics for $\om = \om_-$ of the form  
\be
V \om(\lb, V) = - \frac{4 \pi}{ \ln V} + \frac{(4\pi)^2}{\lb} 
\frac{1}{\ln^2 V} - \frac{(4\pi)^3}{\lb^2} \Big(1 - \lb^2\! \int^{\lb}\! 
\rmd s\,\frac{a(s)}{s^2} \Big) \frac{1}{\ln^3 V} 
+ \rmO\Big(\frac{1}{\ln^4 V}\Big)\,.
\label{gapsol}
\end{equation}

From (\ref{pi01}) one has 
\ba
U_0(\lb , V) \is 2 - \frac{8\pi a(\lb)}{\ln V} + 
\rmO\Big( \frac{1}{\ln^2 V}\Big) \,,
\nonum
U_0(\lb, \infty) \is 2 \,,
\label{binder0}
\ea
in stark contrast to the compact model (as discussed in the conclusions). 

The discussion of the sub-leading order is more involved and is 
best done in Fourier space. We prepare the following auxiliary 
functions
\ba 
\Pi_{st}(p) \is 
\frac{1}{V} \sum_k \frac{1}{(E_k + \om)^s (E_{p-k} + \om)^t}
\,,\,\,\,s,t\ge1\,,
\nonum
\Pi_s(p) \is \Pi_{s1}(p)\,,\,\,\,\,s\ge1\,,
\label{binder1}
\ea
where $\Pi(p) := \Pi_1(p) = \widetilde{\Delta}(p)^{-1}$ is the 
inverse of the Fourier transform of $\Delta(x)$. Further we 
shall need in addition to (\ref{fs}) 
\begin{subeqnarray}
\label{binder2}
f_{st} \is \frac{1}{V^2} \sum_{p,k} \frac{1}
{\Pi(k) (E_p + \om)^s (E_{p-k} +\om)^t}  
=\frac{1}{V}\sum_k\frac{\Pi_{st}(k)}{\Pi(k)}\,.
\end{subeqnarray}
Note
\be
f_{21}=\Pi(0)q\,.
\end{equation}
The expression (\ref{gr4}) for $U_1$ we break up according to 
\ba 
\label{binder3}
U_1 \is U_{11} + U_{12}\,, 
\nonum
U_{11} \is 2 \sigma_1 \sigma_0^{-3} w_0\,,
\\
U_{12} \is - \sigma_0^{-2} w_1 = - \frac{1}{\om^2 V^2} [A + B + C + D]\,.
\nonumber
\ea 
Using $\sigma_1 = 2V \om^{-2}(-q + f_1)$ and $w_0 = 
2V\om^{-2}(V -\om^{-2} \Pi(0)^{-1})$ one has 
\be 
U_{11} = \frac{8}{\om}(q - f_1)\Big(1 - \frac{1}{\om^2 V \Pi(0)}\Big)\,.
\label{binder4}
\end{equation}  
In $U_{12}$ the term $A$ comes from the $\dd w_0/\dd \om$ derivative 
in $w_1$ and is given by 
\be 
A = -2q \om^4 \frac{\dd}{\dd \om} \bigg\{ \frac{2V}{\om^2} \Big( 
V - \frac{1}{\om^2\Pi(0)} \Big) \bigg\} 
= 8q V^2 \Big\{  \om - \frac{2}{V \om \Pi(0)} 
+ \frac{\Pi_{2}(0)}{V \Pi(0)^2} \Big\}\,.  
\label{Aterm}
\end{equation}
The term $B$ corresponds to the $W_2^{(0)} D \Delta D^{-1} W_2^{(0)}$ 
piece in $w_1$ and is given by
\be 
B = -8 \Big(V \om - \frac{2}{\om \Pi(0)} \Big) V f_1 
- \frac{8V}{\Pi(0)^2} f_{31} \,.
\label{Bterm}
\end{equation}  
The term $C$ arises from the $W_2^{(0)} D \Delta \Delta D W_2^{(0)}$
structure and reads
\be 
C = 8 \sum_k \frac{1}{\Pi(k)^2} \Big[ 
\frac{\Pi_2(k)}{\Pi(0)}- \frac{1}{E_k + \om}\Big]^2\,.
\label{Cterm}
\end{equation}  
Finally $D$ comes from the $W_2^{(0)} \Delta W_2^{(0)}$ term 
\be 
\frac{1}{V^2 \om^2} D = - \frac{4}{V \om^2 \Pi(0)}\Big[1 - 2 f_2 +
\frac{1}{\Pi(0)} f_{22}\Big]\,.
\label{Dterm}
\end{equation}   
In the $V \ra \infty$ limit certain terms are superficially divergent, so 
it is best to combine terms where such divergences cancel.
Specifically we rewrite $A+ B$ as 
\ba 
\frac{1}{\om^2 V^2} (A + B) \is  
2 U_{11} + \frac{8}{\om}(q - f_1) 
\Big( \frac{\Pi_2(0)}{V \om \Pi(0)^2} -1 \Big) 
\nonum 
&+& 
\frac{8}{V \om^2 \Pi(0)} \frac{1}{\Pi(0)} (f_1 \Pi_2(0) - f_{31}) \,.
\label{ABterm}
\ea
Further useful combinations to discuss the limit turn out to be  
\begin{subeqnarray}
q - f_1 \is 
\frac{1}{V} \sum_{k} \frac{1}{\Pi(k)} \Big[ 
\frac{\Pi_2(k)}{\Pi(0)}- \frac{1}{E_k + \om}\Big]\,,
\\
\frac{1}{\Pi(0)} f_{22} - 2 f_2 \is 
\frac{1}{V} \sum_k \frac{1}{\Pi(k)} \Big[ 
\frac{\Pi_{22}(k)}{\Pi(0)}- \frac{2}{(E_k + \om)^2}\Big]\,,
\\
f_{31} - \Pi_2(0) f_1 \is \frac{1}{V} \sum_{k} \frac{1}{\Pi(k)} \Big[ 
\Pi_3(k) - \frac{\Pi_2(0)}{E_k + \om} \Big]\,. 
\label{binder5}
\end{subeqnarray}

For the evaluation of the TD limit then mainly the combinations 
in square brackets in (\ref{binder5}) have to be studied, for large 
volumes. A naive replacement of the lattice sums by integrals over 
the Brillouin zone with $\om =0$ would produce infrared divergent 
integrals. The strategy in the following will be to evaluate the 
large volume asymptotics of the functions $\Pi_s,\, s =1,2,3$ and 
$\Pi_{22}$ by repeated insertion of one of the following decompositions
of unity
\begin{subeqnarray}
&& 1 = \frac{1}{E_p + \om} [ E_p - E_{p-k} + 
(E_{p - k} + \om )]\,,
\\
&& 1 = \frac{1}{E_p + 2\om} [ E_p - E_k - E_{p-k} + 
(E_{p - k} + \om ) + (E_k + \om)]\,,
\label{binder7}
\end{subeqnarray}  
until terms corresponding to infrared convergent integrals over the 
Brillouin zone arise. The volume and the momentum dependence of the 
additional pieces picked up in the process (which may diverge as 
$V \ra \infty$) can then be studied analytically. 
For $\Pi_1 = \Pi$ itself only one insertion of (\ref{binder7}b) 
was needed and no divergent piece arose, see (\ref{two6}).
Proceeding similarly we derive the relations
\ba 
\label{binder8}
\Pi_{st}(p) \is \frac{1}{(E_p + 2\om)} \left[X_{st}(p) 
+ \Pi_{(s-1)t}(p) + \Pi_{s(t-1)}(p)\right]\,,\,\,\,\,s,t\ge1\,,
\nonum
X_{st}(p) \is \frac{1}{V} \sum_k  
\frac{E_p - E_k - E_{p-k}}{(E_k + \om)^s(E_{p-k} +\om)^t}\,. 
\ea
Applying it for the case $s=2,t=1$ (and noting 
$\Pi_{s0}(p)=\Pi_{s-1}(0)$) we get 
\be
\label{binder8a}
\Pi_2(p)=\frac{1}{E_p+2\om}\left[ X_{21}(p)+\Pi(p)+\Pi(0)\right].
\end{equation}
This can be rewritten as
\ba 
\label{binder8b}
&& \frac{1}{\Pi(p)} \bigg[ \frac{\Pi_2(p)}{\Pi(0)} - \frac{1}{E_p + \om} \bigg]
\\ 
&& \quad = \frac{1}{ \Pi(0) } \bigg[ \frac{1}{E_p + 2\om} + 
\frac{X_{21}(p)}{(E_p + 2\om)\Pi(p)} \bigg] 
- \frac{\om}{\Pi(p) (E_p + \om) (E_p + 2\om)}\,,
\nonumber
\ea 
from which $q-f_1$ can be evaluated. In a first step one finds 
\ba
&& \Pi(0) (q - f_1 + \om \tilde{f}_2) = 
\frac{1}{V \om} \Big( \frac{\om \Pi_2(0)}{\Pi(0)} -1 \Big)  
\nonum
&& \quad + \frac{1}{V} \sum_{k \neq 0} \frac{1}{E_k + 2\om} + 
\frac{1}{V} \sum_k \frac{J_3(k)}{(E_k + \om)^2} \,.
\label{qf11}
\ea 
Here 
\ba 
\tilde{f_2} \is \frac{1}{V} \sum_{k \neq 0} 
\frac{1}{\Pi(k) (E_k + \om) (E_k + 2\om)}\,,
\nonum
J_3(p) \is \frac{1}{V} \sum_{k \neq 0} 
\frac{E_k - E_p - E_{p-k}}{\Pi(k)(E_{p-k} + \om)(E_k + 2\om)}\,.
\label{qf12}
\ea 
The function $J_3(p)$ has a finite limit which for small $p$ behaves as
\ba 
J_{3,\infty}(p) \is E_p \left[j_3+ \rmO(1/\ln p^2)\right]\,,
\nonum
j_3 \is \int_k \frac{v(k)}{E_k^2}(\cos k_1 -\cos k_2)^2\,. 
\label{qf13}
\ea 
This can be used to show that 
\ba 
\Pi(0) \om f_2 \is \phantom{-}\frac{1}{8\pi} \ln\ln V\, \ln V + 
\frac{q_2}{4\pi} \ln V + \rmO(\ln\ln V) \,,
\nonum
\Pi(0) (q - f_1) \is  -\frac{1}{8\pi} \ln\ln V\, \ln V +\frac{q_1}{4\pi} \ln V 
+ \rmO(\ln\ln V) \,,
\label{qf14}
\ea 
where the constants are related by 
\be 
q_1 + q_2 = j_3\,.
\label{qf15}
\end{equation}   
Indeed, using the fact that 
$E_k \Pi(k)$ has a finite limit which scales like 
$-\frac{1}{2\pi} \ln k^2/T$
for $k^2 \ra 0$, one readily verifies the first equation. Further 
\ba
1 - \frac{\om \Pi_2(0)}{\Pi(0)} \is \rmO\Big( \frac{1}{\ln^2 V} \Big)\,,
\nonum
\tilde{f}_2 + \frac{1}{V \om^2 \Pi(0)} \is f_2 
+ \rmO\Big( \frac{1}{\ln^2 V}\Big)\,,
\nonum
\frac{1}{V} \sum_{k \neq 0} \frac{1}{(E_k + 2\om)} \is 
- \frac{1}{V \om} - \frac{1}{\lb} + \rmO\Big( \frac{1}{\ln V} \Big)\,,
\nonum
\frac{1}{V} \sum_k \frac{J_3(k)}{(E_k + \om)^2} \is 
\frac{1}{4 \pi} j_3 \ln V + \rmO(1)\,. 
\label{qf16}
\ea   
Inserted into (\ref{qf11}) gives
\be 
\Pi(0)[ q - f_1 + \om f_2 ] = j_3 \frac{1}{4 \pi} \ln V + \rmO(\ln\ln V) \,,
\end{equation}  
and hence the first equation in (\ref{qf14}). Combined with 
(\ref{pi01}) we arrive at 
\ba 
&& U_{11} =  16 \pi a(\lb) \frac{\ln \ln V}{\ln V} + 
\rmO\Big( \frac{1}{\ln V} \Big)\,, 
\nonum
&& \frac{8}{\om}(q - f_1) \Big( \frac{\Pi_2(0)}{V \om \Pi(0)^2} -1\Big) 
= - U_{11} +\rmO\Big( \frac{\ln \ln V}{\ln^2 V} \big)\,, 
\label{qf17}
\ea
using 
\be 
1 - \frac{\Pi_{2}(0)}{V \om \Pi(0)^2} = 
1 - \frac{1}{V \om^2 \Pi(0)} + \rmO\Big( \frac{1}{\ln^2 V} \Big) \,.
\end{equation}  
Since $U_1 = U_{11} - (A + B + C + D)/(\om^2 V^2)$ one concludes from 
(\ref{qf17}) and (\ref{ABterm}) that the softly decaying $U_{11}$ terms 
in $U_1$ cancel. 

We proceed with the evaluation of $C$. To this end a more detailed 
evaluation of $X_{21}$ defined in (\ref{binder8}) is needed. 
In a first step one obtains 
\ba
\label{binder9}
& \nspace& (E_p+ \om)^2 X_{21}(p) =  \Big\{ (E_p +\om) \om \Pi(0)\; 
\sum_{\mu} \cos^2\frac{p_{\mu}}{2} + 2 \sum_{\mu} \sin^2 p_{\mu} 
\,\frac{1}{V} \sum_k \frac{(\sin k_1 + \sin k_2)^2}{(E_k + \om)^2}\Big\}
\nonum
&\nspace & \sspace 
+ \Big\{ \frac{E_p + \om}{\lb} \sum_{\mu} \cos^2 \frac{p_{\mu}}{2} +  
J_2(p) + \sum_{\mu} \cos^2\frac{p_{\mu}}{2} \cos p_{\mu} \;
\frac{1}{V}\sum_k \frac{E_k^2}{(E_k + \om)^2} 
\nonum
&\nspace & \sspace 
- 64 \sin^2 \frac{p_1\! -\! p_2}{2}  \sin^2 \frac{p_1\! + \!p_2}{2} \;
\,\frac{1}{V} \sum_k 
\frac{\sin^2 \frac{k_1}{2} \sin^2 \frac{k_2}{2}}{(E_k + \om)^2} \Big\}\,,
\ea
where
\ba
\label{binder10}
&& J_2(p) = \frac{1}{V} \sum_k  
\frac{(E_p - E_k - E_{p-k})(E_p - E_{p-k})^2}{(E_k + \om)^2(E_{p-k} +\om)}\,. 
\nonum
&& -2 \leq  J_{2,\infty}(p) \leq 0 \,,\quad J_{2,\infty}(p) = -2 + j_2\, p^2 + 
\rmO(p^4)\,,
\quad j_2 \approx 0.93\,.
\ea 
As $V \ra \infty$ the first curly bracket diverges logarithmically 
while the second one is convergent. Using 
\ba 
\frac{1}{V} \sum_k \frac{(\sin\frac{k_1}{2} \,\sin\frac{k_2}{2})^2}{(E_k + \om)^2}  
& = & \frac{1}{32 \pi} + \rmO\Big(\frac{1}{V} \Big)\,, 
\nonum
\frac{1}{V} \sum_k \frac{(\sin k_1 + \sin k_2)^2}{(E_k + \om)^2} 
&= & \frac{1}{4\pi} \ln V + a(\lb) - \frac{1}{4} + \frac{1}{4\pi} + 
\rmO\Big(\frac{1}{\ln V} \Big)\,. 
\nonum
\frac{1}{V}\sum_k \frac{E_k^2}{(E_k + \om)^2} &=& 1 
+ \rmO\Big(\frac{1}{V} \Big)\,, 
\ea 
with $a(\lb)$ as in (\ref{pi01}) one finds 
\ba 
\label{binder11}
X_{21}(p) &=& -\frac{1}{E_p^2} (\cos p_1 - \cos p_2)^2 \Big[  
\frac{1}{4\pi} \ln V  + a(\lb) + \frac{1}{2\pi} \Big] 
\\
&+& \frac{1}{E_p^2}\Big[ J_{2,\infty}(p) 
+ \frac{1}{2}\sum_{\mu} (\cos p_{\mu} + \cos2 p_{\mu} ) + 
\frac{1}{2\pi}\sum_{\mu} \sin^2 p_{\mu} \Big] 
+ \rmO\Big(\frac{1}{\ln V} \Big)\,.
\nonumber
\ea
For small momenta this behaves like 
\ba 
\label{binder12}
X_{21}(p) & = &  -\Big[ \frac{1}{4\pi} \ln V  + a(\lb) + \frac{1}{2\pi} 
\Big] 
\Big[ \frac{1}{4} - 
\Big(\frac{p_1}{p_2} + \frac{p_2}{p_1}\Big)^{-2}  + {\rm o}(p_1,p_2) \Big] 
\nonum
&+& \frac{1}{p^2} \Big(j_2  - \frac{5}{4} + \frac{1}{2\pi} + 
{\rm o}(p_1,p_2)\Big) \,.
\ea 

For the evaluation of $C$ it is useful to rewrite (\ref{binder8}) 
for $p \neq 0$ as 
\begin{subeqnarray}
\label{binder13}
\frac{1}{\Pi(p)} \Big[ \frac{\Pi_2(p)}{\Pi(0)} - \frac{1}{E_p + \om} \Big]
\is 
\frac{1}{(E_p + \om) \Pi(0)} \Big[ 
1- \frac{\om \Pi_2(p)}{\Pi(p)} + \frac{X_{21}(p)}{\Pi(p)} \Big]\,,
\\
1- \frac{\om \Pi_2(p)}{\Pi(p)} + \frac{X_{21}(p)}{\Pi(p)} 
\is 
\frac{E_p}{E_p + 2\om} X_2(p) 
+\frac{X_{21}(p)}{\Pi(p)} \frac{E_p + \om}{E_p + 2\om}\,,
\end{subeqnarray}
where 
\be 
X_2(p) := 1 + \frac{\om}{ E_p\Pi(p)} \left(\Pi(p)-\Pi(0)\right)\,.
\end{equation}
In the first term the fact that the numerator is proportional 
to $E_p$ is important for the eventual decay properties of $C$. 
Further $X_2(p)$ is bounded by a function of order 
$\ln V$ in the volume and it has at most logarithmic singularities for 
$p \ra 0$. The latter is consistent with 
\be 
\frac{\om\Pi_2(p)}{\Pi(p)} \sim \frac{\om \Pi(0)}{E_p \Pi(p)} + 
\rmO(1/V)\,,
\end{equation}  
where the $\rmO(1/V)$ piece comes from $\om X_{21}(p)$.

Inserting (\ref{binder13}) into (\ref{Cterm}) gives 
\ba 
C = \frac{8}{\om^2 \Pi(0)^2} \Big( 1 - \frac{\om \Pi_2(0)}{\Pi(0)} \Big)^2 
+ \frac{8V}{\Pi(0)^2} \frac{1}{V} \sum_{k\neq 0}  \frac{1}{(E_k + \om)^2} 
\Big[ 1 - \frac{\om \Pi_2(k)}{\Pi(k)} + \frac{X_{21}(k)}{\Pi(k)} \Big]^2\,.
\label{Cterm2}
\ea 
Using the known behavior of the constituent functions for
large $V$ and small momenta, one can verify a decay of the form% 
\footnote{Here and later on we indicate the form of the sub-leading 
term, without however (in a slight abuse of the ${\rm O}$ symbol) 
presupposing that its coefficient is nonzero.}  
\be 
\frac{C}{\om^2 V^2}  = \rmO\Big(\frac{1}{\ln^2 V}\Big)  +  
\rmO\Big(\frac{1}{\ln^3 V}\Big) \,.
\label{Cterm3}
\end{equation}   

We proceed with the $D$ term, where $\Pi_{22}(p)$ enters. 
A useful representation is 
\begin{subeqnarray} 
\label{Dterm2}
\Pi_{22}(p) \is  \frac{2}{(E_p + 2\om)^2}[\Pi(p) + \Pi(0)] 
+ \frac{6}{(E_p + 2\om)^2} X_{21}(p) 
\nonum
&+& \frac{1}{(E_p + 2 \om)^3} [J_4(p) + 2 J(p) - 2 X_1(p)]\,,
\\
J_4(p) \is \frac{1}{V} \sum_k 
\frac{(E_p - E_{p-k} - E_k)^3}{(E_k + \om)^2 (E_{p-k} + \om)^2}\,,
\\
X_1(p) \is \frac{1}{V} \sum_k \frac{E_p - E_k - E_{p-k}}{(E_k + \om)^2}\,.
\end{subeqnarray}
This can be used to determine the large $V$ behavior of the term 
$D$ in (\ref{Dterm}). We begin by separating the zero mode in the relevant 
combination
\ba 
&& \frac{1}{\Pi(0)} f_{22} - 2 f_2 + 1 =
\Big( 1 - \frac{1}{V \om^2 \Pi(0)} \Big) - 
\frac{1}{V \om^2 \Pi(0)} \Big( 1 - \frac{\om^2 \Pi_{22}(0)}{\Pi(0)} \Big) 
\nonum
&& \quad +  \frac{1}{V} \sum_{k \neq 0} \frac{1}{\Pi(k)} \Big[ 
\frac{\Pi_{22}(k)}{\Pi(0)} - \frac{2}{(E_k + \om)^2} \Big]\,.
\label{Dterm3} 
\ea
On account of 
\ba 
1 - \frac{1}{\om^2 V \Pi(0)} &=& \rmO\Big(\frac{1}{\ln V}\Big) \,,
\nonum
1 - \frac{\om^2 \Pi_{22}(0)}{\Pi(0)} &=& 
1 - \frac{\om \Pi_2(0)}{\Pi(0)} + \rmO\Big( \frac{1}{\ln^3 V} \Big) \,,
\nonumber
\ea 
the zero mode pieces are $\rmO(1/\ln V)$. For the last term on the 
right hand side of (\ref{Dterm3}) we introduce the shorthand 
$S_1 + S_2$. Upon insertion of  (\ref{Dterm2}) we write $S_1$ for the  
part coming from the $\Pi(p) + \Pi(0)$ piece in (\ref{Dterm2}) 
and $S_2$ for the rest, 
\ba
S_1 &=& \frac{2}{\Pi(0)} \frac{1}{V} \sum_{k \neq 0} \frac{1}{(E_k + 
2\om)^2} +\frac{2}{V} \sum_{k \neq 0} 
\frac{1}{\Pi(k)}\Big( \frac{1}{(E_k + 2\om)^2} - 
 \frac{1}{(E_k + \om)^2} \Big)\,. 
\nonum
&=&\rmO\Big(\frac{1}{\ln^2 V} \Big)\,.
\label{Dterm5}
\ea
The term $S_2$ reads
\ba 
\label{Dterm6}
S_2 \is \frac{1}{\Pi(0)} \frac{1}{V} \sum_{k \neq 0} 
\frac{1}{\Pi(k)(E_k + 2\om)^3} [J_4(k) + 2 J(k) - 2X_1(k)] 
\nonum
&+& \frac{6}{\Pi(0)} \frac{1}{V} \sum_{k \neq 0} 
\frac{X_{21}(k)}{\Pi(k)(E_k + 2\om)^2} \,,
\ea 
and is checked to behave as 
\be 
\label{Dterm7}
S_2 = \rmO\Big(\frac{1}{\ln^2 V}\Big)+  \rmO\Big(\frac{1}{\ln^3 V}\Big)\,.
\end{equation}  
Together 
\be 
\frac{D}{\om^2 V^2} =\rmO\Big(\frac{1}{\ln V}\Big) 
+\rmO\Big(\frac{1}{\ln^2 V} \Big)\,.
\label{Dterm8}
\end{equation}  

It remains to consider $A+B$. In view of (\ref{ABterm}) and (\ref{qf17}) 
we know 
\be 
\frac{A+B}{V^2 \om^2} = U_{11} 
+ \rmO\Big( \frac{\ln \ln V }{\ln^2 V} \Big) + 
\frac{8}{V \om^2  \Pi(0)} \frac{1}{\Pi(0)} [ f_1 \Pi_2(0) - f_{31}]\,,
\end{equation}   
so that only $f_{31} - \Pi_2(0) f_1$ is still needed. 
Using (\ref{binder8}) for the case $s=3,t=1$ we obtain
\ba
&& f_{31} - \Pi_2(0) f_1 = \frac{1}{V \om^2 \Pi(0)} 
\Big[ \om^2 \Pi_3(0) - \om \Pi_2(0) \Big]  +
\frac{1}{V} \sum_{k \neq 0} \frac{X_{31}(k)}{\Pi(k)(E_k + 2\om)}
\nonum
&& 
\quad + \frac{1}{V} \sum_{k \neq 0} \frac{\Pi_2(k)}{\Pi(k)(E_k + 2\om)}
- \om \Pi_2(0)\, \frac{1}{V} 
\sum_{k \neq 0} \frac{1}{\Pi(k)(E_k + 2\om)(E_k + \om)}\,. 
\label{binder21}
\ea
Since 
\be
\om^2 \Pi_3(0) - \om \Pi_2(0) = \rmO(V^2)\,, 
\label{binder17}
\end{equation}  
the zero mode piece scales like $\rmO(V/\ln V)$. 
For the second term we observe 
\be
\frac{1}{V} \sum_{k \neq 0} \frac{X_{31}(k)}{\Pi(k)(E_k + 2\om)} = 
\frac{1}{V} \sum_p \frac{J_3(p)}{(E_p + \om)^3} \,,
\end{equation}  
with $J_3$ as in (\ref{qf12}). As a consequence this term 
in (\ref{binder21}) scales like $\rmO(V)$ for large $V$. In the last two terms 
we insert (\ref{binder8a}) to get 
\be 
\frac{1}{V} \sum_{k \neq 0} \frac{1}{(E_k + 2 \om)^2} \Big[ 
1 + \frac{X_{21}(k)}{\Pi(k)} \Big] 
+\frac{1}{V} \sum_{k \neq 0}\frac{1}{ \Pi(k) (E_k + 2 \om)^2} 
\left[\Pi(0)-\om\Pi_2(0)\frac{E_k+2\om}{E_k+\om}\right]\,. 
\label{binder23}
\end{equation}  
The very first term is $\rmO(V \ln V)$, the one involving $X_{21}(p)$
is $\rmO(V/\ln V)$, and  the last one is $\rmO(V \ln \ln V)$. 
Together 
\be 
f_{31} - \Pi_2(0) f_1 = \rmO(V \ln V) + \rmO(V \ln \ln V)\,.
\label{binder24}
\end{equation}  
For $A+B$ this results in 
\be 
\frac{1}{ V^2 \om^2}(A + B) = U_{11} + \rmO\Big( \frac{1}{\ln V} \Big) + 
\rmO\Big( \frac{\ln \ln V}{\ln^2 V} \Big) \,.
\label{binder25}
\end{equation}  

Combining (\ref{binder3}) with (\ref{binder25}), (\ref{Cterm3}) and 
(\ref{Dterm8}) we arrive at the conclusion:
\ba 
U_1(\lb , V) \is \rmO\Big( \frac{1}{\ln V} \Big) + 
\rmO\Big( \frac{\ln \ln V}{\ln^2 V} \Big) + \rmO\Big( 
\frac{1}{\ln^2 V} \Big)\,,
\nonum
U_1(\lb ,\infty) \is 0\,.
\label{binder26}
\ea
For the TD limit of the full Binder cumulant the result 
(\ref{binder26}) amounts to
\be 
U(\lb, \infty) = \frac{2}{N\!+\!1}  + 
\rmO\Big(\frac{1}{(N\!+\!1)^3} \Big)\,. 
\label{binder}
\end{equation}  
This result will be backed by Monte-Carlo simulations 
in Subsection~5.2. Potential implications for 
``criticality'' and ``triviality'' of the theories are 
discussed in the conclusions.

%%%%%%%%%%%%%%%%%%%%%%%%%%%%%%%%%%%%%%%%%%%%%%%%%%%%%%%%%%%%%%%%%%
\newsubsection{Direct evaluation of lattice sums} 

Both as a check on the previous analysis and in order to have finite 
volume data to compare Monte-Carlo data with, we also evaluated the 
lattice sums defining $U_1$ and several other quantities numerically 
up to $L=1024$. Since $\rmO(L^4)$ terms have to be summed and both very 
small (e.g.~$\omega$) and very large numbers (e.g.~$\Pi(0)$) 
enter high precision is needed. The summations were performed 
to 96 bit (26 significant figures) accuracy using the publicly available 
arbitrary precision MPFR library ({\tt www.mpfr.org}) and for moderate 
$L$ also with Mathematica.

The results were found to vary with $\lb$ such that for 
smaller $\lb$ the presumed large $V$ asymptotics sets in later. 
Below we present the results for $\lb =3$; qualitatively 
those for other $\lb$ values are similar. Due to the predicted  
occurrence of very slowly varying terms (e.g.~of $\ln\ln V/\ln V$ 
type) one cannot expect that the genuine large $V$ asymptotics 
can be unambiguously probed by direct summation. Nevertheless 
two or three parameter fits of the $L \leq 1024$ sums to the 
expected decay form are generally convincing. Table 1 summarizes 
results for $\om$, 
$\Pi(0)$ and some slowly varying quantities entering $U_{11}$.     
Here $U_{11}$ is defined in Eq.~(\ref{binder4}), $q-f_1$ 
is evaluated directly from (\ref{qdef}), (\ref{fs}) and 
from (\ref{binder5}a), $f_2$ is defined in (\ref{fs}). For 
example the leading asymptotics $\om \Pi(0) \sim 
- \frac{1}{4\pi} \ln V$ and the coefficients in 
\be 
- f_2 \sim \frac{1}{2} \ln \ln V \sim \frac{1}{\om} (q-f_1)\,,
\end{equation}
come out well in fits to the data.

\vspace{8mm} 

\begin{table}[htb]
\label{qf1f2data}
\centering

\begin{tabular}{|c|c|c|c|c|c|}
\hline
$L$ & $10^6\,\om$  & $(q-f_1)/\om$ & $f_2$ & $10^{-6}\,\Pi(0)$ \\[0.5mm] 
\hline
 64  &  $-$233.007495171 & $-$0.25538202207 & 0.42884117188 & 0.00451330538 \\[0.5mm]
 128 &  $-$52.7199137013 & $-$0.23716419150 & 0.37935605329 & 0.02202558370 \\[0.5mm] 
 256 &  $-$12.0362079518 & $-$0.21754473457 & 0.33427610710 & 0.10558882967  \\[0.5mm]
 512 &  $-$2.76867400687 & $-$0.19723910189 & 0.29289048252 & 0.49868470422  \\[0.5mm] 
 768 &  $-$1.17557207628 & $-$0.18522986021 & 0.27016715051 & 1.22915930848 \\[0.5mm]
 1024 & $-$0.64094863000 & $-$0.17669596989 & 0.25464412160 & 2.32558377444 \\[0.5mm]
\hline
 \end{tabular}
\caption{\small Quantities entering $U_{11}$ for $\lb =3$; all given digits 
are significant.}
\end{table}
\vspace{4mm}

Table 2 presents results 
for the terms used in the breakup of $U_1$, see Eq.~(\ref{binder3}),
and the final result for $U_1$. The column for $A+B$ 
again illustrates the need for high precision, as 
individually $A$ and $B$ are $2-5$ orders of 
magnitudes larger that their sum. It also highlights that the analytical 
evaluation of the large $V$ asymptotics is crucial. Even at $L =1024$ the 
normalized $A+B$ contribution is still increasing. On account of 
(\ref{binder25}) the decay of the combination $(A+B)/(\om V)^2 - U_{11}$ 
should be faster. Indeed these data have a maximum at around $L =300$ and 
then decay monotonically in a way fitted well by the predicted functional 
form.

\begin{table}[htb]
\label{U2data}
\centering

\begin{tabular}{|c|c|c|c|c|c|}
\hline
\\[-4.5mm]
$L$ & $10^2\,U_{11}$  & $10^3\,(A+B)/(V\om)^2$ & 
$10^3\,C/(V \om)^2$ & $10^3\,D/(V \om)^2$ & $U_1$\\[0.5mm] 
\hline\\[-4.5mm]
 64  & $-$0.748704 & $-$4.65746  &  9.13703 & 0.387281 & $-$0.012353890 \\[0.5mm]
 128 & $-$0.565668  & $-$2.24328 & 7.25786 & 0.275149 & $-$0.010946420 \\[0.5mm] 
 256 & $-$0.431162 & $-$0.73430 & 5.92122 & 0.189879 & $-$0.009688426 \\[0.5mm]
 512 & $-$0.330143 &  \phantom{$-$}0.23478 & 4.92648 & 0.132507 &  $-$0.008595209\\[0.5mm] 
 768 & $-$0.282636 &  \phantom{$-$}0.63551 & 4.45707 & 0.108460 &  $-$0.008027416\\[0.5mm]
 1024 & $-$0.253114 & \phantom{$-$}0.86471 & 4.16358 & 0.094567 &  $-$0.007654026 \\[1mm]
\hline
 \end{tabular}
\caption{\small Quantities contributing to $U_1$ for $\lb =3$; 
all given digits are significant.}
\end{table}

\vspace{4mm}

Finally we present a fit of the $U_1$ data to the 
predicted decay form in (\ref{binder26}).

%%%%%%%%%%%%%%%%%%%%%%%%%%%%%%%%%%
\begin{figure}[htb]
\leavevmode

\hspace{2cm}
\epsfxsize=10cm
\epsfysize=7cm
\epsfbox{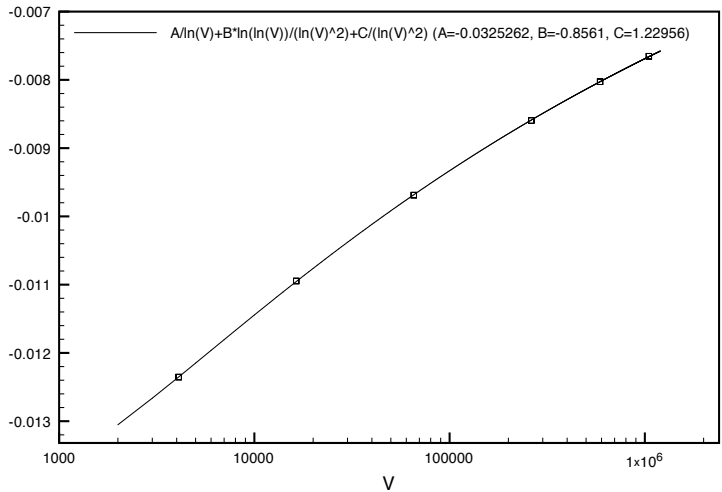}
\vspace{-2mm}

\caption{\small  $U_1$ vs $V$. 
Fit to $\rmO(\ln \ln V/\ln^2 V) + \rmO(1/\ln V) + \rmO(1/\ln^2 V)$.}
\label{U2plot}
\end{figure}
%%%%%%%%%%%%%%%%%%%%%%%%%%%%%%%%%%%  

%%%%%%%%%%%%%%%%%%%%%%%%%%%%%%%%%%%%%%%%%%%%%%%%%%%%%%%%%%%%%%%%%%%%%%%%
\newsubsection{MC results for $U$} 

Since the large $N$ expansion is only an asymptotic expansion 
the higher order coefficients in (\ref{gr4}) are not bound to be small,
even in finite volume. At any given $N$ the truncated series 
could in principle misrepresent the exact $U(\lb,V)$. In order to
preclude this possibility we estimated $U(\lb,V)$ via 
Monte-Carlo simulations. 

We have chosen to simulate a ${\rm SO}(1,8)$ theory at $\lambda =3$ 
on lattices of linear dimensions $L=32, 64, 128, 256, 384$. The simulations 
were performed in a fixed spin gauge (the spin at the origin
was held fixed). The variable spins were updated by a Metropolis procedure
tuned to achieve a roughly 50\% acceptance rate. Equilibration and
autocorrelation times for various observables have an enormous range: 
non-gauge-invariant observables in particular (e.g. $\bra n^{0}\ket$) 
require extremely long runs, and on larger lattices fail to reach 
equilibrium even after billions of Monte-Carlo sweeps. The
situation is much better for gauge-invariant observables, such as
$\Sigma\! \cdot\! \Sigma$ entering $U_1$, see (\ref{gr1}). The fluctuations 
in this latter quantity determine the Binder cumulant, and are typically 
stable after a few million sweeps. Results for the quantity 
$\frac{2}{N+1}-U$ versus lattice size are shown in Fig.~2. This
quantity is {\em not} monotonic, but reaches a maximum near $L=64$ and 
then decreases quite rapidly. The decrease appears to be faster 
than the log-type decay found for $U_{1}$ in Subsection~5.1, 
suggesting that the termwise large $V$ asymptotics of 
the large $N$ series (when formally treated as convergent) 
sums to a power-like large volume decay.

%%%%%%%%%%%%%%%%%%%%%%%%%%%%%%%%%%
\begin{figure}[htb]
\leavevmode

\hspace{2.3cm}
\epsfxsize=10cm
\epsfysize=7cm
\epsfbox{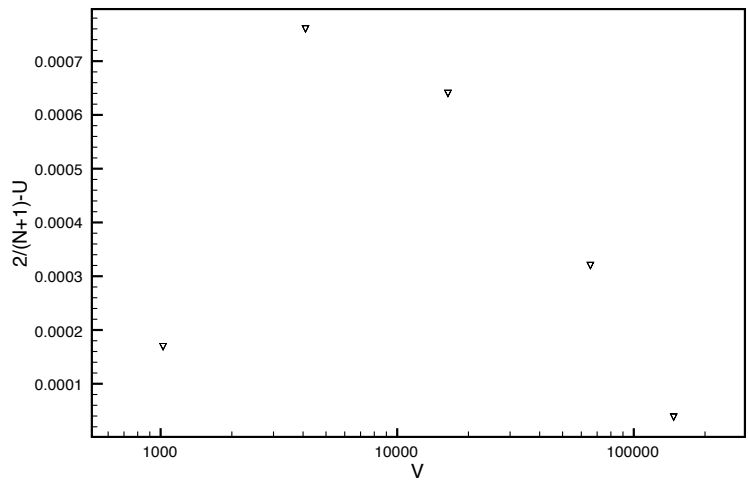}
\vspace{-3mm}

\caption{\small  Monte-Carlo results for $\frac{2}{N+1} -U$ vs $V$}. 
\label{U2MCplot}
\end{figure}
%%%%%%%%%%%%%%%%%%%%%%%%%%%%%%%%%%%  

In  summary, the numerical evidence suggests that the 
large $N$ contributions to the Binder cumulant beyond 
leading order (i.e.~$U_{0}(\lambda, V)$) may indeed vanish in 
the thermodynamic limit.

%%%%%%%%%%%%%%%%%%%%%%%%%%%%%%%%%%%%%%%%%%%%%%%%%%%%%%%%%%%%%%%%%%%%%%%%%%

\newsection{Conclusions}

Noncompact ${\rm SO}(1,N)$ sigma-models are expected to 
be massless in contrast to their compact counterparts. 
The infrared problem therefore is nontrivial, especially in 
dimension $d=2$, and the 
goal of the present paper has been to gain computational 
control over the limit of vanishing infrared regulator. 
The large $N$ expansion is well suited for this; 
in a lattice formulation  the dynamically generated gap 
is negative and serves as a coupling dependent infrared regulator 
which vanishes in the limit of infinite lattice size. 
The cancellation of infrared divergences has been 
demonstrated in $d=2$ by explicit computation of a number 
of physically interesting quantities defined in terms of 
invariant correlation functions: the spin and current two-point 
functions as well as the Binder cumulant, all to next to leading
order. A complementary result is \cite{SZ} where 
a noninvariant observable was shown to have a finite 
thermodynamic limit in $d \geq 3$ beyond large $N$. 
In $d=2$ we expect that a `large $N$' counterpart 
of David's theorem \cite{David} can be established, showing 
that infrared divergences cancel termwise in the large $N$ 
expansion of invariant correlation functions to all orders.

To discuss our result for the Binder cumulant let us 
first recall the situation in the compact model. 
In the notation of Subsection~3.4 one has there  
$1/\Pi_+(0) \sim 4\pi \om_+$, in the thermodynamic (and continuum) 
limit, so that $VU = 8\pi/[(N\!+\!1)\om_+]$.
Taking $\xi = 1/\sqrt{\om_+}$ as the 
definition of the correlation length, this gives the 
familiar result for the renormalized coupling 
${\rm g}_r = VU/\xi^2 = 8\pi/(N\!+\!1)$, to leading 
order. See also ref.~\cite{campostrini}
for a direct continuum computation to sub-leading order, with 
the result $(N\!+\!1){\rm g}_r = 
8\pi[ 1 - 0.602033/(N\!+\!1) + \rmO(1/(N\!+\!1)^2)]$.   

In the noncompact model the zero momentum limits of 
invariant correlation functions are expected to 
diverge in the thermodynamic limit. Indeed, mostly 
this reflects the fact that they are increasing 
functions of the lattice distance (recall 
$n_x \cdot n_y \geq 1$, always). Our results of Section~5
suggest however that the ratio entering $U$ is finite, 
independent of $\lb$, and very close to $2/(N\!+\!1)$. 

One can view this result as a manifestation of a ``concentration 
of measure'' phenomenon. For the 1D lattice model with $L$ sites 
it was shown in \cite{hchain} that the functional measure has 
support mostly on configurations boosted by an amount increasing 
at least powerlike with $L$. In the thermodynamic limit the 
measure (or mean) is therefore concentrated `at infinity', i.e.~in 
the disc model of the hyperbolic geometry at the boundary of 
the disc. Though not proven in dimensions $d >1$ it is very plausible 
that a similar concentration phenomenon will hold for the 
$d$-dimensional functional measures. Indeed our result on the 
Binder cumulant can be put into this context: First note 
that in terms of the normalized average spins $\sigma^a := 
\Sigma^a/\sqrt{\Sigma \cdot \Sigma}$, with $\Sigma^a = \sum_x n^a_x$, 
one has    
\be 
{\rm Var}(\sigma^2) := \bra (\sigma\cdot \sigma  - 
\bra \sigma \cdot \sigma\ket)^2 \ket = \frac{2}{N+1} -U \geq 0\,.
\end{equation}      
In the thermodynamic limit ${\rm Var}(\sigma^2)$ has been 
argued to vanish, which is natural if the components of 
$\sigma^a$ are typically very large rendering the relative 
fluctuations ensuring $\bra \sigma \cdot \sigma \ket =1$ 
and  $\bra (\sigma \cdot \sigma)^2 \ket \approx 1$, negligible. 
Indeed to leading order of the large $N$ expansion 
one finds $\bra \sigma^0 \ket \sim \sqrt{\ln V}$. Note 
that the constant $2/(N\!+\!1)$ can be interpreted as the value 
of $U$ in a constant configuration and that the indefinite dot 
product is crucial here.

Alternatively $2/(N+1) -U$ is given by the ratio of 
the susceptibilities defined from the partially connected
4-point and 2-point functions $W_2$ and $W_1$, respectively. 
Both diverge as $V \ra \infty$, but the ratios entering the 
large $N$ expansion of $U$ (viz $w_1/\sigma_0^2$, $w_2/\sigma_0^2$, 
see Eqs.~(\ref{UW}) -- (\ref{u01})) vanish in the 
thermodynamic limit. This is compatible with a genuine 
factorization of the 4-point function but does not entail it. 
(Since the connected four point function 
$\bra n_{x_1}\! \cdot\! n_{y_1} n_{x_2}\! \cdot 
\!n_{y_2}\ket_c$ entering (\ref{gr1}) does not take into 
account the nonzero one-point functions, the fact that 
it must be non-zero is only indirectly relevant for this.)

Concerning local quantities, the analysis of the 
TD limit for the subleading term of the spin-two point function 
in Subsection 4.1 does not suggest the existence of a nontrivial 
limit as the UV cutoff is removed. Positive bare couplings 
are required for the large $N$ series to be an asymptotic 
expansion, in which case a naturally defined renormalized coupling 
vanishes as the UV cutoff is removed. The situation should be 
similar for the two-point function of the Noether current. 

Together, our results may be taken as an indication for ``triviality''
of the theory in the sector comprising ${\rm SO}(1,N)$ invariant 
observables. If corroborated beyond the large $N$ expansion this 
would be of significance in a number of other contexts, e.g.~for
a class of Kaluza-Klein theories or for the widely studied systems 
with ${\rm AdS}_5\times S_5$ target spaces. The focus on invariant 
observables is certainly natural from the viewpoint of the 
compact models. In the context of the Osterwalder-Schrader 
reconstruction \cite{hchain,DNS}, however, invariant correlators are not 
ideal for the non-compact systems, 
and the situation may well be be different when noninvariant observables 
are considered.

%%%%%%%%%%%%%%%%%%%%%%%%%%%%%%%%%%%%%%%%%%%%%%%%%%%%%%%%%%%%%%%%%%%%%%%%%%%%%%
\bigskip 

{\it Acknowledgements:} We wish to thank E.~Seiler for many discussions, 
and M. L\"{u}scher for correspondence. The research of A.D.~is supported 
in part by the National Science Foundation under grant PHY-0554660.

%%%%%%%%%%%%%%%%%%%%%%%%%%%%%%%%%%%%%%%%%%%%%%%%%%%%%%%%%%%%%%%%%%%%%%%%%%

\newpage
%\newappendix
\setcounter{section}{0}
\newappendix{Leading and next-to-leading order SD equations}

Here we tabulate the first few of the hierarchy of Schwinger-Dyson  
equations for the large $N$ coefficients $W_r^{(s)}$, $r + s >1$. 
They can be obtained e.g.~by first converting (\ref{sd1}) 
into a system of equations for the exact $W_r$ and then 
inserting the large $N$ ansatz (\ref{c5a}). The ensued  
recursive structure is summarized in Fig.~1. 

The leading 2--point function (\ref{k10}) satisfies
\be
\left[\pm\Delta_x-\lambda\right] W_{\pm,1}^{(0)}(x,y)
\mp\left[\lambda W_{\pm,1}^{(0)}(x,y)+1\right]\Delta_x 
W_{\pm,1}^{(0)}(x,z)|_{z=x}=1-\delta_{xy}\,.
\label{sdk10}
\end{equation}
In the next order we have
\ba
&&\pm\left[\Delta_x-\om_\pm\right] W_{\pm,1}^{(1)}(x,y)
-\lambda D_\pm(x,y)\Delta_x W_{\pm,1}^{(1)}(x,z)|_{z=x}
\nonumber\\
&&=-\left(1-\delta_{xy}\right)\left[\lambda W_{\pm,1}^{(0)}(x,y)+1\right]
\pm\lambda\Delta_x W_{\pm,2}^{(0)}(x,z;z,y)|_{z=x}\,,
\label{sdk11}
\ea
where we have used the solution (\ref{k10}) to the leading order 
equation (\ref{sdk10}) to simplify some terms.

We see that to solve (\ref{sdk11}) we first need to solve the
equation for the leading order 4--point function: 
\ba
\label{sdk20}
&&\pm\left(\Delta_{x_1}-\om_\pm\right)W_{\pm,2}^{(0)}(x_1,y_1;x_2,y_2)
-\lambda W_{\pm,1}^{(0)}(x_1-y_1)\Delta_{x_1}
W_{\pm,2}^{(0)}(x_1,z;x_2,y_2)|_{z=x_1}
\nonum
&&=\mp\left[\delta_{x_1x_2}D_\pm(y_1-y_2)+\delta_{x_1y_2}D_\pm(y_1-x_2)\right]
\nonum
&&+\lambda\left[\delta_{x_1x_2}+\delta_{x_1y_2}\right]
D_\pm(x_1-y_1)D_\pm(x_2-y_2)\,.
\ea
The solution for $W_{\pm,2}^{(0)}$ is given in (\ref{k20}), from
which can verify that $W_{\pm, 1}^{(1)}$ in (\ref{k11}) 
solves (\ref{sdk11}).  

In the next order the equation for the 4--point function is
\ba
\label{sdk22}
&&\pm\left[\Delta_{x_1}-\om_\pm\right]W_{\pm,2}^{(1)}(x_1,y_1;x_2,y_2)
-\lambda D_\pm(x_1-y_1)\Delta_{x_1}W_{\pm,2}^{(1)}(x_1,z;x_2,y_2)|_{z=x_1}
\nonum
&&=\pm\lambda\Delta_{x_1}W_{\pm,3}^{(0)}(x_1,u;v,y_1;x_2,y_2)|_{u=v=x_1}
\\
&&\pm\lambda W_{\pm,2}^{(0)}(x_1,y_1;x_2,y_2)
\Delta_{x_1}W_{\pm,1}^{(1)}(x_1,z)|_{z=x_1}
\nonum
&&\pm\lambda W_{\pm,1}^{(1)}(x_1,y_1)
\Delta_{x_1}W_{\pm,2}^{(0)}(x_1,z;x_2,y_2)|_{z=x_1}
\nonum
&&-\lambda W_{\pm,2}^{(0)}(x_1,y_1;x_2,y_2)   
-\delta_{x_1x_2}W_{\pm,1}^{(1)}(y_1,y_2) 
-\delta_{x_1y_2}W_{\pm,1}^{(1)}(x_2,y_1) 
\nonum
&&+\lambda\left(\delta_{x_1x_2}+\delta_{x_1y_2}\right)
\Bigl\{ W_{\pm,2}^{(0)}(x_1,y_1;x_2,y_2)
\nonum
&&
\pm D_\pm(x_2-y_2)W_{\pm,1}^{(1)}(x_1,y_1) 
\pm D_\pm(x_1-y_1)W_{\pm,1}^{(1)}(x_2,y_2)\Bigr\}\,.
\nonumber
\ea
One sees the pattern summarized in Fig.~1 emerging, 
in that the solution of (\ref{sdk22}) requires knowledge of the 
leading order 6--point function. The latter satisfies the equation:
\ba
\label{sk23}
&&\pm\left(\Delta_{x_1}-\om_\pm\right)
W_{\pm,3}^{(0)}(x_1,y_1;x_2,y_2;x_3,y_3)
\nonum
&&-\lambda 
D_\pm(x_1-y_1)\Delta_{x_1}W_{\pm,3}^{(0)}(x_1,z;x_2,y_2;x_3,y_3)|_{z=x_1}
\nonum
&&=\pm W_{\pm,2}^{(0)}(x_1,y_1;x_2,y_2)\Delta_{x_1}
W_{\pm,2}^{(0)}(x_1,z;x_3,y_3)|_{z=x_1}
\\
&&\phantom{=}\pm W_{\pm,2}^{(0)}(x_1,y_1;x_3,y_3)\Delta_{x_1}
W_{\pm,2}^{(0)}(x_1,z;x_2,y_2)|_{z=x_1}
\nonum
&&-\delta_{x_1x_2}W_{\pm,2}^{(0)}(y_1,y_2;x_3,y_3)
  -\delta_{x_1x_3}W_{\pm,2}^{(0)}(y_1,y_3;x_2,y_2)
\nonum
&&-\delta_{x_1y_2}W_{\pm,2}^{(0)}(y_1,x_2;x_3,y_3)
  -\delta_{x_1y_3}W_{\pm,2}^{(0)}(y_1,x_3;x_2,y_2)
\nonum
&&\pm\lambda\left[\delta_{x_1x_2}+\delta_{x_1y_2}\right]
\left[D_\pm(x_1-y_1)W_{\pm,2}^{(0)}(x_2,y_2;x_3,y_3)
     +D_\pm(x_2-y_2)W_{\pm,2}^{(0)}(x_1,y_1;x_3,y_3)\right]
\nonum
&&\pm\lambda\left[\delta_{x_1x_3}+\delta_{x_1y_3}\right]
\left[D_\pm(x_1-y_1)W_{\pm,2}^{(0)}(x_2,y_2;x_3,y_3)
+     D_\pm(x_3-y_3)W_{\pm,2}^{(0)}(x_1,y_1;x_2,y_2)\right]\,,
\nonumber
\ea
where all the functions on the rhs are known from solutions of 
(\ref{sdk10}) and (\ref{sdk20}). The solution is simply given by
\ba
\label{sk24}
&& W_{\pm,3}^{(0)}(x_1,y_1;x_2,y_2;x_3,y_3)=
\pm\sum_{w_1,w_2,w_3}\sum_{z_1,z_2,z_3}
\nonumber\\
&& \quad \times W_{\pm,2}^{(0)}(x_1,y_1;w_1,z_1)
  W_{\pm,2}^{(0)}(x_2,y_2;w_2,z_2)
  W_{\pm,2}^{(0)}(x_3,y_3;w_3,z_3)
\nonum
&&\quad \times D_\pm^{-1}(w_1-z_2)D_\pm^{-1}(w_2-z_3)D_\pm^{-1}(w_3-z_1)\,.
\ea
Using (\ref{sk24}) one can verify that $W_{\pm,2}^{(1)}$ as given in 
(\ref{k21}) solves (\ref{sdk22}).

%%%%%%%%%%%%%%%%%%%%%%%%%%%%%%%%%%%%%%%%%%%%%%%%%%%%%%%%%%%%%%%%%%%%%%%%%%%%%

\setcounter{section}{1}
\newpage
\newappendix{Continuum limit behavior of $j(p)$}

In this appendix we consider the small $p$ behavior of the function 
$j(p)$ entering the spin two-point function (\ref{two5}) to subleading 
order. We have
\be
J_{\Pi\infty}(p)-J_{\Pi\infty}(0)=E_pj_1(p)+\sum_{\mu}\sin p_\mu 
\,j_{2;\mu}(p)+\sum_\mu \hat{p}_\mu^2 \,j_{3;\mu}(p)\,,
\end{equation}
with
\ba
j_1(p)&=&-\int_k\frac{1}{E_{k-p}}\left[v(k)-v(p)\right]\,,
\nonum
j_{2;\mu}(p)&=&\int_k\frac{\sin k_\mu}{E_k}
\left[v(p-k)-v(p+k)\right]\,,
\\[2mm]
j_{3;\mu}(p)&=&\frac14\int_k\frac{\hat{k}_\mu^2}{E_k}
\left[v(p-k)+v(p+k)\right]\,,
\nonumber
\ea
where $v(p)$ is defined in (\ref{piinf}) and $\hat{k}_{\mu} = 
2 \sin \frac{k_{\mu}}{2}$, as usual. 
Note first $j_{3;\mu}(p)$ is non-singular at $p=0$:
\be
j_{3;\mu}(0)=-\frac14 J_{\Pi\infty}(0)\,.
\end{equation}
Next
\ba
j_{2;\mu}(p)&=&-2\int_k\frac{\sin(k-p)_\mu}{E_{k-p}}[v(k)-v(p)]
\nonum
&=&-\sin p_\mu[2j_1(p)+j_{4;\mu}(p)]-2\cos p_\mu\, j_{5;\mu}(p)\,,
\ea
with
\ba
j_{4;\mu}(p)&=&\int_k\frac{\hat{k}_\mu^2}{E_{k-p}}[v(k)-v(p)]\,,
\nonum
j_{5;\mu}(p)&=&\int_k\frac{\sin k_\mu}{E_{k-p}}[v(k)-v(p)]
= 2\sum_\nu\sin p_\nu \,j_{6;\mu\nu}(p)\,,
\\[2mm]
j_{6;\mu\nu}(p) &=&\int_k\frac{\sin k_\mu\sin k_\nu}
{E_{k-p}E_{k+p}}\left[v(k)-v(p)\right]\,.
\nonumber
\ea
Noting
\be
j_{4;\mu}(0)=-\frac12 J_{\Pi\infty}(0)\,,
\end{equation}
we have for small $p^2$:
\be
j(p)\sim -j_1(p)-4\frac{\sin p_\mu\sin p_\nu}{E_p}j_{6;\mu\nu}(p)
+\frac14 J_{\Pi\infty}(0)+\rmO(1/\ln p^2)\,.
\end{equation}

From (\ref{two8}) we have
\be
v(p) = \alpha(p)+\rmO(p^2)\,,\sspace 
\alpha(p) := -\frac{2\pi}{\ln(p^2/T)}\,,
\end{equation}
with $T$ defined in (\ref{defT}). We now consider two corresponding integrals 
\ba
\tilde{j}_1(p)&=&-\int_k^{\infty}
\frac{\theta(c^2-k^2)}{(k-p)^2}
\left[\alpha(k)-\alpha(p)\right]\,,
\nonum
\tilde{j}_{6;\mu\nu}(p)&=&\int_k^{\infty}
\frac{\theta(c^2-k^2)k_\mu k_\nu}{(k-p)^2(k+p)^2}
[\alpha(k)-\alpha(p)]\,,
\ea
where $\int_k^{\infty}$ denotes $\int_{-\infty}^\infty\rmd^2 k/(2\pi)^2$
and $c$ is a momentum cutoff $T>c^2>p^2$. These give the leading
small $p^2$ contribution because
\ba
j_1(p)-\tilde{j}_1(p)&=&v_1(c)+\rmO(1/\ln p^2)\,,
\nonum
j_{6;\mu\nu}(p)-\tilde{j}_{6;\mu\nu}(p)
&=&-\frac12\delta_{\mu\nu}\left[v_1(c)+\frac14 v_2\right]+\rmO(1/\ln p^2)\,,
\label{B10}
\ea
with
\ba
v_1(c)&=&-\int_k^\infty
\Big[\frac{v(k)}{E_k}\prod_\mu\theta(\pi-|k_\mu|)
-\frac{\alpha(k)}{k^2}\theta(c^2-k^2)\Big]\,,
\nonum
v_2&=&\int_k\frac{\sum_\mu\hat{k}_\mu^4 \,v(k)}{E_k^2}\,.
\ea

First using
\be
\int_0^{2\pi}\rmd\phi\,\frac{1}{(t+\cos\phi)}=\frac{2\pi}{\sqrt{t^2-1}}\,,\,
\quad t^2>1\,,
\end{equation}
we can do the angular integrations in the $\tilde{j}$ functions, setting 
without loss of generality $p\mapsto (p,0)$:
\ba
\tilde{j}_1(p)&=&\frac12\int_0^{c^2} \rmd x\frac{1}{|x-p^2|}
\left[\frac{1}{\ln(x/T)}-\frac{1}{\ln(p^2/T)}\right]\,,
\nonum
\tilde{j}_{6;00}(p)&=&\frac{1}{8p^2}\int_0^{c^2} \rmd x
\left[1-\frac{x+p^2}{|x-p^2|}\right]
\left[\frac{1}{\ln(x/T)}-\frac{1}{\ln(p^2/T)}\right]\,.
\ea

Noting $c^2>p^2$ we obtain
\be
\tilde{j}_1(p)
=-\frac{1}{2\ln(p^2/T)}\left[S_1(p^2,T)+S_2(p^2,T)+S_3(p^2,c^2,T)\right]\,,
\label{B14}
\end{equation}
with
\ba
S_1(p^2,T)&=&\int_0^1\rmd y\,
\frac{\ln y}{(1-y)\ln(p^2y/T)}\,,
\nonum
S_2(p^2,T)&=&\int_0^1\rmd y\,
\frac{\ln(1+y)}{y\ln[(1+y)p^2/T]}\,,
\\[2mm]
S_3(p^2,c^2,T)&=&\int_1^{c^2/p^2-1}\rmd y\,
\frac{\ln(1+y)}{y\ln[(1+y)p^2/T]}\,.
\nonumber
\ea
Now for small $p^2$
\ba
S_1(p^2,T)&\sim&\frac{1}{\ln(p^2/T)}\sum_{n=0}^\infty 
s_1^{(n)}\frac{1}{[\ln(p^2/T)]^n}\,,
\nonum
S_2(p^2,T)&\sim&\frac{1}{\ln(p^2/T)}\sum_{n=0}^\infty  
s_2^{(n)}\frac{1}{[\ln(p^2/T)]^n}\,,
\ea
with
\ba
s_1^{(n)}&=&(-1)^n\int_0^1\rmd y\,\frac{[\ln y]^{n+1}}{(1-y)}\,,
\nonum
s_2^{(n)}&=&(-1)^n\int_0^1\rmd y\,\frac{[\ln(1+y)]^{n+1}}{y}\,,
\ea
giving $s_1^{(0)}=-\frac{\pi^2}{6}\,,s_2^{(0)}=\frac{\pi^2}{12}\,,\dots$
Next
\be
S_3(p^2,c^2,T)=S_4(p^2,c^2,T)+S_5(p^2,c^2,T)\,,\quad c^2 <T\,.
\end{equation}
Here 
\ba
& \nspace & S_4(p^2,c^2,T) = \int_1^{c^2/p^2-1}\rmd y\,
\frac{\ln(1+y)}{(1+y)\ln[(1+y)p^2/T]}
= \int_2^{c^2/p^2}\rmd z\,\frac{\ln(z)}{z\ln(zp^2/T)}
\nonum
&\nspace & = \int_{\ln 2}^{\ln(c^2/p^2)}\rmd x\,\frac{x}{x+\ln(p^2/T)}
= -\ln(2p^2/c^2)+\ln(p^2/T)\ln\left(\frac{\ln(2p^2/T)}{\ln(c^2/T)}\right)\,,
\ea
and
\ba
S_5(p^2,c^2,T)&=&\int_1^{c^2/p^2-1}\rmd y\,
\frac{\ln(1+y)}{y(1+y)\ln[(1+y)p^2/T]}
\nonum
&\sim&\frac{1}{\ln(p^2/T)}\sum_{n=0}^\infty s_5^{(n)}
\frac{1}{[\ln(p^2/T)]^n}\,,
\ea
with 
\be
s_5^{(n)}=(-1)^n\int_1^\infty\rmd y\,\frac{[\ln(1+y)]^{n+1}}{y(1+y)}\,,
\end{equation}
giving $s_5^{(0)}=\frac{\pi^2}{12}+\frac12(\ln2)^2\,\dots$. So
by (\ref{B10}) and (\ref{B14}) 
\ba
j_1(p) &\sim & -\frac12\ln\left(-\ln(2p^2/T)\right)+g_1
+\rmO\left([\ln(p^2/T)]^{-1}\right)
\nonum
g_1 &= & v_1(c)+\frac12\ln\left(-\ln(c^2/T)\right)+\frac12\,,
\ea
which is independent of $c$. Similarly 
\be
\tilde{j}_{6;00}(p)=\frac{1}{4\ln(p^2/T)}
\left[S_6(p^2,T)+S_2(p^2,T)+S_3(p^2,c^2,T)\right]\,,
\end{equation}
with
\be
S_6(p^2,T)=\int_0^1 \rmd y\,\frac{y\ln y}{(1-y)\ln(p^2y/T)}
\sim\frac{1}{\ln(p^2/T)}\sum_{n=0}^\infty 
s_6^{(n)}\frac{1}{[\ln(p^2/T)]^n}\,,
\end{equation}
where
\be
s_6^{(n)}=(-1)^n\int_0^1\rmd y\,\frac{y[\ln y]^{n+1}}{(1-y)}\,,
\end{equation}
giving $s_6^{(0)}=1-\frac{\pi^2}{6}\,,\dots$.  

Putting all the results together we obtain (\ref{smallplatt})
with 
\be
g_2=g_1+\frac12 v_2+\frac14 J_{\Pi\infty}(0)\,.
\end{equation}

%%%%%%%%%%%%%%%%%%%%%%%%%%%%%%%%%%%%%%%%%%%%%%%%%%%%%%%%%%%%%%%%%%%%%

\setcounter{section}{2}
\newpage
\newappendix{Large $N$ with two auxiliary fields}  

The results for the large $N$ expanded correlation functions 
in the noncompact model have in Section~3 been obtained via the 
large $N$ correspondence summarized in Subsection~2.2. Direct large $N$ 
computations in the noncompact model can be based on the following 
generating functional \cite{EMP} 
\ba
\label{wdnc}
\exp W^d_-[H] \is  \exp\Big\{-\frac{1}{2} \sum_{x,y} H_{xy} \Big\} 
\,\cN\! \int\! \prod_{x \neq x_0} \rmd\alpha_x 
\exp\Big\{ - (N\!+\!1) S_-[\alpha,H]\Big\}\,,
\nonum
S_-[\alpha,H] \is \frac{1}{2} 
\Tr \ln \widehat{A} + i \sum_{x \neq x_0} \alpha_x 
-\frac{1}{2\lb} (\widetilde{A}^{-1})_{x_0x_0}^{-1} \,,
\nonum
A_{xy} \is - \Delta_{xy} + 2 i \lb \alpha_x\delta_{xy} + 
\frac{\lb}{N+1} H_{xy} 
= \widetilde{A}_{xy} + 2 i \lb \delta_{xy} \delta_{x x_0} \alpha_{x_0}\,.
\end{eqnarray} 
Here $\widehat{A}$ is the matrix obtained by deleting the 
$x_0$-th row and column of $A$ or $\widetilde{A}$. 
The formal expansion based on (\ref{wdnc}) is not a valid saddle 
point expansion but it does produce the correct expansion coefficients 
and is related to its counterpart $W^d_+[H]$ in the compact model 
by the involution $\alpha_x \mapsto -\alpha_x$, $\lambda \mapsto 
-\lambda$. The functional (\ref{wdnc}) thus provides a 
simple heuristic way to understand the large $N$ correspondence.
In contrast to the compact model, however, $W_-^d[H]$ is {\it not} 
equivalent to the original generating functional $W_-[H]$.

Here we outline how (\ref{wdnc}) can formally be obtained  
from the formulation of the large $N$ expansion with two auxiliary 
fields introduced in \cite{DNS}. We begin by dualizing the 
`spatial'  spin components $\vec{n}_x,\,x \in \Lambda$,  
as one would do in the compact model. Indeed, the $N$ spatial 
components $\vec{n}_x,\,x \neq x_0$, enter 
(\ref{wnc}) with the `good' sign; their `dualization' gives 
\ba 
\label{a1}
&& \exp W_-[H] = \exp\Big\{-\frac{1}{2} \sum_{x,y} H_{xy} \Big\}
\int \prod_x \rmd n_x^0 \delta(n_{x_0}^0 -1)
\\
&& \quad \times\! \int\! \prod_{x \neq x_0} \rmd\alpha_x 
\exp\Big\{ - \frac{N}{2} \Tr \ln \widehat{A} - i (N+1) \sum_{x\neq x_0} 
\alpha_x \Big\} 
\exp\Big\{ + \frac{N+1}{2\lb} \sum_{x,y} n_x^0 
\widetilde{A}_{xy} n_y^0\Big\}\,.
\nonumber
\end{eqnarray}
Here $\hat{A}$ arises due to the constrained Gaussian integration. 
Note the small 
but crucial differences to the compact model \cite{EMP}: 
only $N$ copies of $\Tr \ln \widehat{A}$ occur so far and the sign of the  
$\sum_{x \neq x_0} \alpha_x$ term is flipped, as is the sign 
of the $H_{xy}$ term in $A_{xy}$. Most importantly the 
kinetic term in the last exponential has the wrong sign,
which is why one cannot naively interchange the order
of the integrations over $n_x^0$ and $\alpha_x,\,x \neq x_0$. 
To proceed we assume that in a large $N$ expansion the
replacement 
\be 
\int \prod_x \rmd n_x^0 \delta(n_{x_0}^0 -1) = 
(-)^{|\Lambda|/2} \int \prod_x \rmd\eta_x \delta(\eta_{x_0}) \,,\quad 
n_x^0 = \bar{n}_x + i \eta_x\,,
\label{a2}
\end{equation}
is legitimate, for certain ``saddle point'' configurations $\bar{n}_x$,
with $\bar{n}_{x_0} =1$. With this replacement the kinetic term acquires 
the good sign. After the additional re-routing $\alpha_x = -i \om_x/(2\lb) + 
\xi_x$ the saddle point conditions $\dd S/\dd \eta_x = \dd S/\dd \xi_x =0$
lead to 
\be 
\bar{n}_x^2 = 1 + \lb \widehat{D}_{xx}\,,\quad 
(\Delta \bar{n})_x = \om_x \bar{n}_x \,,\quad x \neq x_0\,.
\label{a3}
\end{equation}
(See Eq.~(5.18) of \cite{DNS}, with $\om_x = 2 \lb \bar{\alpha}_x$, and 
correcting the sign in the second formula).  
Here $D_{xy} = (M^{-1})_{xy}$ with $M_{xy} := -\Delta_{xy} + 
\delta_{xy} \om_x$. Since $-\Delta$ is a positive operator it follows 
from the second equation in (\ref{a3}) that the position dependent 
$\omega_x$ must be predominantly negative: $0\leq  - \sum_x 
\bar{n}_x (\Delta \bar{n})_x = - \sum_x \om_x \bar{n}_x^2$.

To leading order the spin two-point function is given by \cite{DNS} 
\ba 
\label{a4}
&& \bra n_x \cdot n_y \ket_{f.s.} = \bar{n}_x \bar{n}_y - 
\lb \widehat{D}_{x,y}\,,
\nonum
&& \widehat{D}_{x,y} := D_{x,y} - \frac{D_{x,x_0} D_{y,x_0}}{D_{x_0,x_0}} \,,
\ea 
where we write momentarily $\bra \;\; \ket_{f.s}$ for the average 
computed with the fixed spin measure in (\ref{wnc}). The quantity 
$\bar{n}_x$ then is the nonzero expectation value $\bra n_x^0 \ket_{f.s.}$ 
to leading order in $1/(N+1)$. 

To make contact to the gap equation (\ref{gap}) in Subsection~3.1 
we now first replace (\ref{a3}) by a simpler gap equation 
with constant $\om$ and $\bar{n}$,
\be
\bar{n}^2 - \lb D'(0) = 1,\quad \bar{n}^2 \om = - \frac{\lb}{V}\,,
\quad \mbox{with}\quad 
D'(x) := \frac{1}{V} \sum_{p\neq 0} \frac{\rme^{ip \cdot x}}{E(p) + \om}\,. 
\label{a5}
\end{equation}
These are the saddle point conditions arising from a translation invariant 
gauge fixing of the functional integral (see Eq.~(5.6) of \cite{DNS}) and 
imply $-\lb D(0) =1$, in accordance with (\ref{gap}). 
Given a solution $\om, \bar{n}$ of (\ref{a5}) we claim that 
\be 
\bar{n}_x := - \lb D(x-x_0)\,,\quad 
\om_x := \om + \lb \delta_{x, x_0}\,,
\label{a6}
\end{equation}
is a solution of (\ref{a3}). The equation $(\Delta \bar{n})_x = 
(\omega + \lb \delta_{x, x_0}) \bar{n}_x$ is checked using 
$- \lb D(0) =1$. To verify the first equation in (\ref{a5}) 
it suffices to observe that 
\be 
\widehat{D}_{x,y} = D(x-y) - D(x-x_0) D(y-x_0)/D(0)\,.
\label{a7}
\end{equation}
This can be seen as follows: suppose that 
invertible matrices $M$ and $\widetilde{M}$ are related by  
$M_{xy} = \widetilde{M}_{xy} - c \delta_{xy} \delta_{x_0 x}$. 
Then the inverse of $M$ is related to the inverse of $\widetilde{M}$ by 
\be 
(M^{-1})_{xy} = (\widetilde{M}^{-1})_{xy} + 
\frac{c}{ 1 - c (\widetilde{M}^{-1})_{x_0x_0}} 
(\widetilde{M}^{-1})_{x x_0}(\widetilde{M}^{-1})_{y x_0}\,.
\label{d6}
\end{equation} 
Applied to $M_{xy} = - \Delta_{xy} + \om_x \delta_{xy} = \widetilde{M}_{xy} 
+ \lb \delta_{xy} \delta_{x,x_0}$, for $\om_x = \om + \lb 
\delta_{x,x_0}$, without yet assuming the gap equation (\ref{a5}), 
this gives first 
\be 
D_{xy}= D(x-y) - \frac{\lb}{1 + \lb D(0)} D(x-x_0) D(y-x_0)\,,
\label{a8}
\end{equation} 
and then (\ref{a7}) from (\ref{a4}). In particular 
no pole occurs for the quantities in (\ref{a7}) at $\lb D(0) = -1$.   
Using (\ref{a7}) and $\lb D(0) =-1$, it follows $1 + \lb \widehat{D}_{x,x} 
= \lb^2 D(x-x_0)^2 = \bar{n}_x^2$, while the sign in (\ref{a6}) is fixed by 
$-\lb D(x) \geq 1$.

Inserting (\ref{a6}) into (\ref{a4}) we arrive at 
\be 
\bra n_x \cdot n_y \ket_{\rm f.s.} = - \lb D(x-y) = \bar{n}^2 - \lb D'(x-y) 
= \bra n_x \cdot n_y \ket_{\rm trans} \,,
\label{a9}
\end{equation}
where the right hand side coincides with the spin two-point function 
computed to leading order in $1/(N+1)$ in the translation 
invariant gauge \cite{DNS} and with that of Subsection~3.1.

In summary, the leading order results with two auxiliary fields 
and the two gauge fixings considered (fixed spin and translation 
invariant gauge) are related via (\ref{a6}). Both saddle point 
equations (\ref{a3}) and (\ref{a5}) imply the version $-\lb D(0) =1$ 
used here, but in addition provide the interpretation of 
$\bar{n}_x = \bra n_x^0 \ket_{\rm f.s.}$ and 
$\bar{n} = \bra n_x^0 \ket_{\rm trans}$, as the averages of 
$n_x^0$ with respect to the respective gauge-fixed functional measures.    
Note that $\bar{n}_x$ approaches $\bar{n}^2$ as $|x-x_0|$ becomes 
large and that $\sum_x \bar{n}_x = V \bar{n}^2$. Although the 
invariant two-point functions coincide to leading order in the 
two gauges, the results for the noninvariant quantity 
$\bra n_x^0\ket$ are very different, $\bra n_x^0\ket_{f.s} = 
\bra n_x^0\ket_{\rm trans}^2$.   

Equipped with this interpretation of $\bar{n}_x$ we return to (\ref{a1}). 
Subject to the assumption (\ref{a2}) one can proceed by interchanging 
the order of integrations, which results in the Gaussian 
\ba 
\label{a10}
&& \int \prod_x \rmd\eta_x \,\delta(\eta_{x_0}) 
\exp\Big\{\! - \!\frac{N+1}{2\lb} \sum_{x,y} \eta_x \widetilde{A}_{xy} \eta_y 
+ \frac{N+1}{\lb} \sum_x \eta_x \sum_y i \widetilde{A}_{xy} \bar{n}_y\Big\}
\nonum
&& \quad = {\rm Const}\, (\det \widehat{A})^{-1/2}
\exp\Big\{ \frac{N+1}{2\lb} \Big[ \bar{n}_{x_0}^2 
(\widetilde{A}^{-1})_{x_0x_0}^{-1} - \sum_{x,y} \bar{n}_x \widetilde{A}_{xy} 
\bar{n}_y\Big] \Big\}\,.
\ea 
Using also $\bar{n}_{x_0} =1$ and substituting back into (\ref{a1}) 
we arrive at (\ref{wdnc}).

%%%%%%%%%%%%%%%%%%%%%%%%%%%%%%%%%%%%%%%%%%%%%%%%%%%%%%%%%%%%%%%%%%%%%%%%%%%%%

%%%%%%%%%%%%%%%%%%%%%%%%%%%%%%%%%%%%%%%%%%%%%%%%%%%%%%%%%%%%%%%%%%%%%%%%%%%%%
\end{document}